# Surface Restructuring of Nickel Sulfide Generates Optimally-Coordinated Active Sites for ORR Catalysis

Bing Yan,[1][†] Dilip Krishnamurthy,[1][‡] Christopher H. Hendon,[†] Siddharth Deshpande,[‡] Yogesh Surendranath,*[†][2] Venkatasubramanian Viswanathan,*[‡]

[†]Department of Chemistry, Massachusetts Institute of Technology, 77 Massachusetts Avenue, Cambridge, Massachusetts 02139-4307, United States
[‡]Department of Mechanical Engineering, Carnegie Mellon University, 5000 Forbes Avenue, Pittsburgh, Pennsylvania 15213, United States

**First-row transition metal oxides and chalcogenides have been found to rival or exceed the performance of precious metal-based catalysts for the interconversion of water and $O_2$, central reactions that underlie renewable electricity storage and utilization. However, the high lability of the first-row transition metal ions leads to surface dynamics under the conditions of catalysis and results in active site structures distinct from those expected by surface termination of the bulk lattice. While these surface transformations have been well-characterized on many metal oxides, the surface dynamics of heavier chalcogenides under electrocatalytic conditions are largely unknown. We recently reported that the heazlewoodite $Ni_3S_2$ bulk phase supports efficient ORR catalysis under benign aqueous conditions and exhibits excellent tolerance to electrolyte anions such as phosphate which poison Pt. Herein, we combine electrochemistry, surface spectroscopy and high resolution microscopy to characterize the surface dynamics of $Ni_3S_2$ under ORR catalytic conditions. We show that $Ni_3S_2$ undergoes self-limiting oxidative surface restructuring to form an approximately 2 nm amorphous surface film conformally coating the $Ni_3S_2$ crystallites. The surface film has a nominal NiS stoichiometry and is highly active for ORR catalysis. Using DFT simulations we show that, to a first approximation, the catalytic activity of nickel sulfides is determined by the Ni-S coordination numbers at surface exposed sites through a simple geometric descriptor. In particular, we find that the surface sites formed dynamically on the surface of amorphous NiS during surface restructuring provide an optimal energetic landscape for ORR catalysis. This work provides a systematic framework for characterizing the rich surface chemistry of metal-chalcogenides and provides principles for the development of structure-energy-activity descriptors leading to a broader understanding of electrocatalysis mediated by amorphous materials.**

The interconversion of water and oxygen is a central chemistry underlying the storage of renewable electricity in energy-dense chemical bonds(Lewis and Nocera, 2006). The oxidation of $H_2O$ to $O_2$ is the efficiency limiting half reaction for the splitting of water to generate $H_2$ fuel, whereas the reduction of $O_2$ to $H_2O$ is the efficiency limiting cathode reaction in low temperature fuel cells(Katsounaros et al., 2014). Platinum group metals and their corresponding oxides and chalcogenides are well-known catalysts for these reactions(Matsumoto and Sato, 1986)-(Gasteiger et al., 2005), but recent studies have uncovered a diversity of earth abundant first-row transition metal oxides(Kanan and Nocera, 2008)-(Long et al., 2014) and chalcogenides(Gao et al., 2012)·(Gao et al., 2013) that, depending on the reaction conditions, rival the activity of their precious metal analogs.

Unlike their precious metal congeners, first-row transition metal ions are labile,(Helm and Merbach, 1999, 2005) and as a result, the surfaces of these materials are expected to be more dynamic under the conditions of catalysis. Indeed, the remarkable propensity for first-row transition metal oxides to reconstruct and in many cases amorphize under the condition of the oxygen evolution reaction (OER) has been well documented.(Lee et al., 2012; Trotochaud et al., 2014; Yeo and Bell, 2011) Understanding these surface dynamics is critical for developing rigorous structure-function correlations, because the surface phase rather than the bulk pre-catalyst carries out the desired reaction. However, compared to the rich contemporary understanding of surface phase transformations under OER conditions, there exists a paucity of information about the analogous dynamics that give rise to active surface phases for the oxygen reduction reaction (ORR). In part, this is due to the limited number of heterogeneous ORR catalysts containing first-row transition metal ions. Pt alloys of Ni and Co present a notable exception. For these alloys, the labile first-row transition metal component undergoes oxidative dissolution, leaving behind Pt-rich skins that display enhanced ORR activity.(Toda, 1999)·(Stamenkovic et al., 2007) Despite these precedents, the surface phase dynamics of Pt-free ORR catalysts remain almost entirely unexplored despite the expectation that they would be even more susceptible to surface reconstruction under catalytic conditions.

---

[1] These authors contributed equally.

[2] Lead Contact

First-row late transition metal chalcogenides (LTMCs) containing Fe, Co, and Ni have been identified as a potent class of catalysts for the ORR.(Baresel et al., 1974)·(Behret et al., 1975) The best performing LTMCs exhibit ORR onset potentials in the range of 0.7-0.8 V vs the reversible hydrogen electrode (RHE), requiring only ~0.1 V greater overpotential than Pt. In addition to being composed entirely of low-cost, earth-abundant elements, LTMCs display excellent tolerance to crossover of fuels such as methanol and common fuel contaminants such as CO, making them particularly promising as low cost materials for next generation low-temperature fuel cell technologies.(Feng et al., 2011) Despite their attractive qualities, the systematic improvement of the performance of these materials has been impeded by a lack of coherent structure-function correlations that can be used as the basis for rational catalyst design. Indeed, there remains a great deal of ambiguity about the surface structures that are responsible for catalysis because of the paucity of investigations that examine surface phase dynamics under catalytically relevant conditions.(Susac et al., 2006)

Recently, we reported that the heazlewoodite, $Ni_3S_2$, bulk phase supports efficient ORR catalysis under benign neutral pH conditions.(Falkowski et al., 2015) The material displays superior performance relative to the corresponding first-row transition metal oxides with an ORR onset potential of 0.80 V and, unlike Pt, is remarkably tolerant to poisoning by electrolyte ions such as phosphate. These properties, combined with the known tolerance of LTMCs to common liquid fuels, such as formate, make this material particularly attractive for low cost, membrane-free fuel cells.(Cheng et al., 2006; Yan et al., 2017; Zhao et al., 2005) In order to enable the systematic development of improved earth-abundant ORR catalysts, we now probe the surface phase dynamics of this material to identify the active surface structures. We find that the surface layers of $Ni_3S_2$ undergo a self-limiting oxidative reconstruction under ORR conditions to generate a conformal ~2 nm thick amorphous nickel sulfide surface layer with an approximate stoichiometry of Ni:S 1:1. We show via independent synthesis that this amorphous surface phase can account for the high catalytic activity of $Ni_3S_2$ for the ORR under neutral pH conditions.

In order to probe the diverse surface chemistry that emerges upon reconstruction, we performed density functional theory calculations for ORR on several crystalline and amorphous nickel sulfide phases. Our analysis indicates that to a first approximation, the key activity descriptor for ORR, the free energy of adsorbed OH*, is determined by the local bonding environment of surface exposed Ni sites and dominated by their nearest-neighbor sulfur coordination number. We find that surface Ni sites that have three sulfur neighbors possess optimal activity and such sites exist at the surfaces of several nickel sulfide phases. Furthermore, the same geometric descriptor is found to hold for amorphous layers with a NiS stoichiometry. We apply these observations to show that the restructured amorphous surface layer hosts optimally-coordinated Ni sites primed for efficient ORR catalysis. This work provides a systematic framework for characterizing the rich surface chemistry on metal-chalcogenides and provides principles for the development of structure-activity descriptors. These principles and the descriptor-based approach allow for the rapid identification of classes of materials competent for the ORR and can accelerate the discovery of earth-abundant catalysts.

**Results and discussion**

The Ni-S phase diagram consists of six stoichiometric nickel sulfide phases which range in composition from $Ni_3S_2$ to $NiS_2$.(Bugajski et al., 2004) The heazlewoodite end member phase, $Ni_3S_2$, examined in this study, crystallizes in the trigonal space group R32 with each Ni bound to four S and four Ni nearest neighbor atoms (Figure S1a). The Ni-Ni bonds in $Ni_3S_2$ are 2.50 Å, only slightly longer than the corresponding bond length of 2.44 Å observed in Ni metal. We synthesized $Ni_3S_2$ by combining Ni and S powder in a stoichiometric ratio, followed by heating under vacuum in a sealed quartz tube at 600 °C for 20 hours and then at 800 °C for another 4 hours.(Tare and Wagner, 1983) Powder X-ray diffraction (PXRD) (Figure S1b) of this material reveals peaks corresponding only to heazlewoodite Ni$_3$S$_2$(Fleet, 1977) evincing the formation of a phase-pure sample. TEM and SEM reveal average crystallite dimensions of several hundred nanometers (Figure S2).

The surfaces of the as-prepared Ni$_3$S$_2$ samples are structurally similar to the bulk lattice. In line with literature reports, X-ray photoelectron spectroscopy (XPS) of the as synthesized Ni$_3$S$_2$ shows a Ni 2p$_{3/2}$ peak at 852.7 eV, and a S 2p peak at 162.5 eV(Buckley and Woods, 1991) (Figure 1a and 1b, black) in a 3:2 atomic ratio, along with an adventitious O peak of low intensity at 531.8 eV. In addition, high resolution transmission electron microscopy (TEM) images of Ni$_3$S$_2$ particles uniformly display lattice fringes that extend to the edge of each crystallite (Figure S3a and b). Together, these data indicate that the surfaces of Ni$_3$S$_2$ particles remain crystalline and are not subject to significant deterioration prior to the electrochemical investigations detailed below.

The surfaces of Ni$_3$S$_2$ undergo self-limiting oxidative phase conversion under electrochemical polarization. Freshly-prepared Ni$_3$S$_2$ particulates were pressed onto gold electrodes and were examined by cyclic voltammetry (CV) in N$_2$-saturated 1 M sodium phosphate (NaP$_i$) electrolyte, pH 7. Consistent with Ni$_3$S$_2$ having a formal Ni-valency of 4/3, we observed a low open circuit potential of 0.30 V (all potentials are reported versus the reversible hydrogen electrode, RHE). Scanning negative of this value reveals no discernable CV features up to 0.00 V. However, on the first positive going scan, we observe a large irreversible oxidative wave with a peak potential of $E_{p,a}$ = 0.47 V (Figure 1d). Notably, the peak of this oxidative feature resides ~0.30 V negative of the onset potentials for ORR on the Ni$_3$S$_2$ material. Upon scanning positive of the anodic peak, the current rapidly declines, reaching a background level with no additional oxidative features up until 0.9 V. Importantly, this oxidative peak is not accompanied by any reductive features on the subsequent negative going scan and the magnitude of the oxidative peak rapidly declines in subsequent scans and is not detectable by the seventh CV cycle of a freshly prepared electrode. Together, the electrochemical data establish that Ni$_3$S$_2$ surfaces undergo irreversible self-limiting oxidation under electrochemical polarization to the potentials necessary for ORR catalysis.

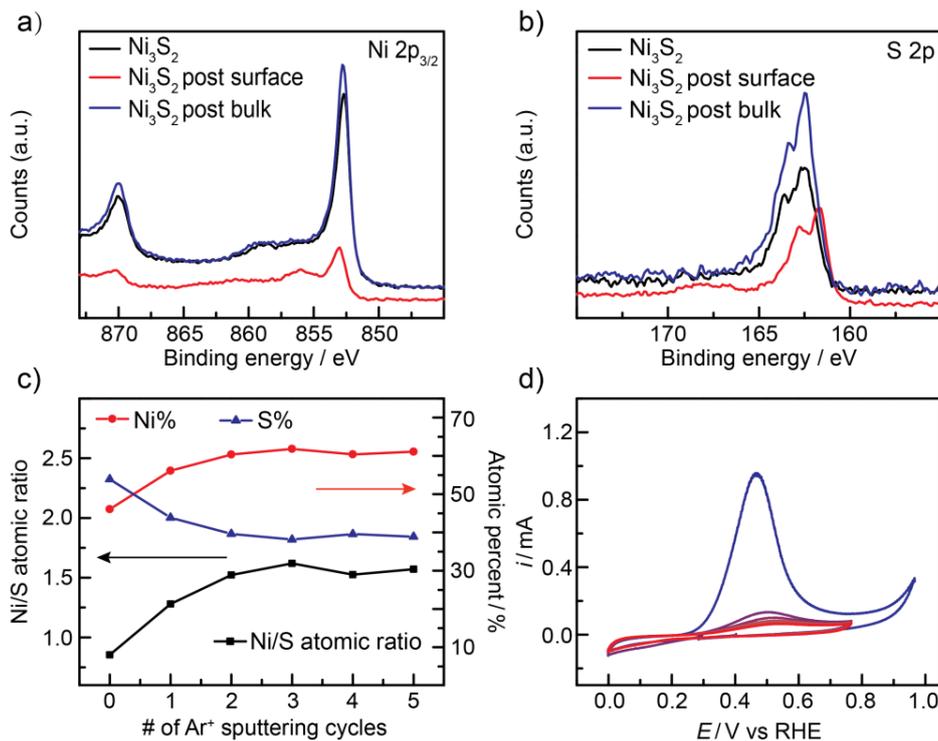

**Figure 1.** High resolution XPS of (a) Ni 2p$_{3/2}$ and (b) S 2p regions. As prepared Ni$_3$S$_2$, post-oxidation surface, and bulk Ni$_3$S$_2$ are shown in black, red, and blue respectively. (c) XPS-derived Ni and S atomic ratios and percentages as a function of the number of Ar$^+$ sputtering cycles. (d) Initial CV scans of Ni$_3$S$_2$ recorded in N$_2$-saturated 1 M sodium phosphate electrolyte, pH 7, at a scan rate of 5 mV s$^{-1}$ and rotation rate of 2000 rpm.

XPS of the CV-cycled Ni$_3$S$_2$ electrode reveals a reconstructed surface with a reduced Ni stoichiometry. On the surface, the Ni 2p$_{3/2}$ peak shifts slightly to higher binding energy (BE) from 852.7 to 853.0 eV and we observe a small peak at 856.0 eV corresponding to a minority surface population of Ni(OH)$_2$ (Figure 1a, red).(Buckley and Woods, 1991),(Casella et al., 1999) Additionally, the S 2p peak shifts from 162.5 to 161.7 eV, similar to the peak position observed for NiS (Figure 1b, red).(Buckley and Woods, 1991),(Han et al., 2003) Additionally, following surface reconstruction, we observe a Ni:S atomic ratio of 0.9:1.0 (Figure 1c). We also observe an O 1s peak at 531.60 eV (Figure S4, blue) attributed to a minority surface population of $Ni(OH)_2$ or surface adsorbed hydroxide species.(Casella et al., 1999) Following argon ion ($Ar^+$) etching of the surface, the peak positions of Ni and S both return to the values typical of $Ni_3S_2$ and the O peak disappears (Figure 1a and 1b, blue; Figure S4, red and black). Additionally, the Ni:S atomic ratio increases to 1.3:1.0 and 1.5:1.0 after one and two cycles of $Ar^+$ sputtering respectively. The value stabilizes at ~1.5:1.0, in line with the bulk stoichiometry of $Ni_3S_2$ (Figure 1c). Together the XPS results reveal that the surfaces of $Ni_3S_2$ undergo oxidative reconstruction upon CV cycling in neutral electrolyte to form a new surface phase with an approximate empirical formula of NiS.

High resolution TEM (HRTEM) images suggest that the surface phase conversion is accompanied by morphology changes. As the above described solid-state synthesis produced particles too large (a few hundred nanometers) to discern changes in nanoscale morphology, we prepared $Ni_3S_2$ nanoparticles following a procedure adapted from the literature(Chi et al., 2012) and used these samples for HRTEM investigations. Similar to the solid-state prepared samples, the as-prepared nanoparticles display uniform $Ni_3S_2$ lattice fringes extending to the edge of each crystallite (Figure 2a, Figure S3c and S3d). These nanoparticles were affixed to a TEM grid and subjected to potential cycling as described above until the oxidative peak declined to the background level. Following this electrochemical treatment, we observe a thin amorphous "shell" conformally coating each particle (Figure 2b and S5). By sampling the shell thickness across ten particles, we obtained a histogram showing an average shell thickness of $1.6 \pm 0.6$ nm (Figure S6a and b). To gain more insight into this self-limiting surface phase transformation, we performed chronoamperometry (CA) in $O_2$-saturated neutral electrolyte on the freshly prepared $Ni_3S_2$ nanoparticles. At 0.71 V, we observed an initial anodic current corresponding to oxidative restructuring, which rapidly decays to a steady state cathodic value corresponding to ORR catalysis (Figure S6c). The histograms of the amorphous surface layer thickness after 23 s (immediately prior to the onset of net cathodic current), 2 min, and 5 min of CA polarization (Figure S6d, e and f) reveal an average amorphous layer thickness of $1.5 \pm 0.7$, $2.0 \pm 0.9$ and $1.9 \pm 0.4$ nm respectively. Prolonged polarization beyond these time points does not lead to any further changes in the thickness of the amorphous layer, consistent with the self-limiting nature of the oxidative restructuring. To further probe this self-limiting surface restructuring, we performed pH dependence studies by recording CVs on freshly prepared $Ni_3S_2$ electrodes in pH 4.68, 7.02, 8.37 and 9.87 sodium phosphate electrolyte, respectively (Figure S7, details are discussed in the Supporting Information). The experimental data are in line with the Pourbaix diagram (Figure S8) we constructed from the thermodynamic data of known nickel compounds, further substantiating the formation of amorphous NiS (*a*-NiS) upon electrochemical cycling at neutral pH. Together, the data indicate that the surface reconstruction of $Ni_3S_2$ generates an amorphous ~2.0 nm *a*-NiS surface layer that serves to passivate the $Ni_3S_2$ host from further oxidation.

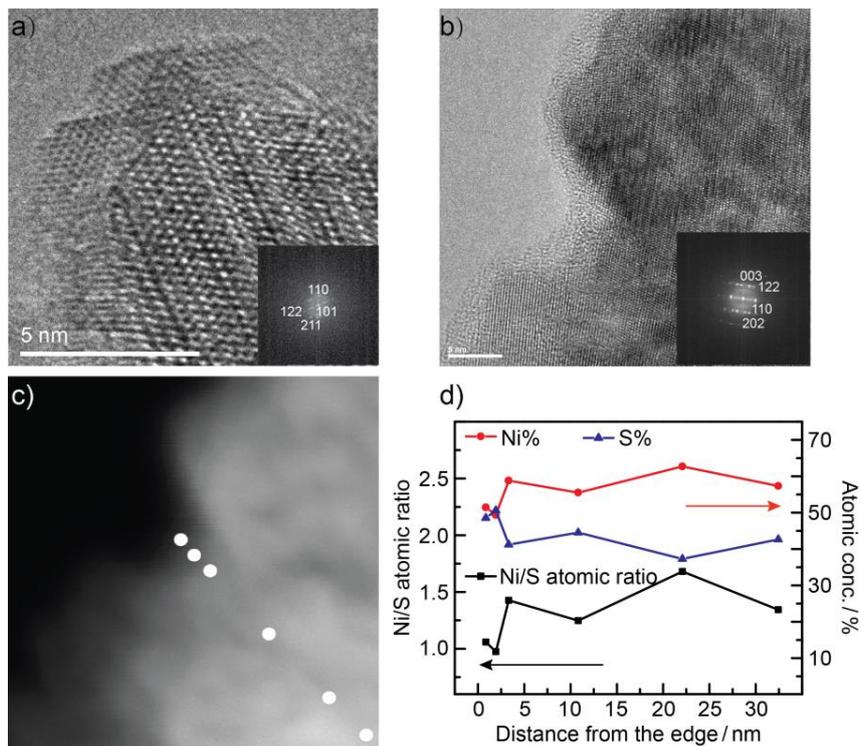

**Figure 2.** TEM images of (a) as-prepared and (b) post-electrolysis $Ni_3S_2$ nanoparticles. Fast-Fourier transforms are shows in the insets of (a) and (b) and confirm the crystalline nature of the bulk material. Elemental composition determined from energy-dispersive X-ray spectroscopy (EDS) at a series of spots along a line from the crystallite edge to the bulk (c) reveals the variation in Ni:S composition across the crystallite (d).

Energy-dispersive X-ray spectroscopy (EDS) under scanning transmission electron microscopy (STEM) mode further confirms a Ni:S composition change in the thin surface amorphous layer. EDS data were collected at a variety of points extending from the edge of the nanoparticle to its interior (Figure 2c) and the corresponding Ni:S atomic ratio is plotted in Figure 2d. At the edge of the particle, where the amorphous layer is present, the Ni:S atomic ratio is ~1.0:1.0 and this ratio rapidly increases to ~1.5:1.0 as the scan proceeds to the interior of the $Ni_3S_2$ particle, reflecting the bulk stoichiometry. These TEM data, combined with the electrochemical and XPS data described above, indicate that the surfaces of $Ni_3S_2$ undergo oxidative restructuring at the potentials necessary for ORR catalysis, forming an amorphous ~2 nm layer with a nominal NiS stoichiometry.

To determine whether the amorphous surface layer is the active phase, we compared the ORR activity of surface-restructured $Ni_3S_2$, with independently prepared crystalline and *a*-NiS. We electrodeposited NiS on an Au RDE substrate using a method adapted from previous reports.(Jana et al., 2014),(Sun et al., 2013) The as-prepared NiS film was amorphous as judged by the absence of peaks in grazing incidence defractograms (Figure S10a). Following thermal annealing at 250 ˚C under $N_2$ for three hours, the crystallinity of the film is improved and grazing incident XRD (GIXD) and XPS evince the formation of crystalline NiS surface phases (Figure S10b, c and d). The ORR activity of these materials were assessed in $O_2$-saturated 1 M sodium phosphate (NaP$_i$), pH 7. The as-prepared $Ni_3S_2$ electrode undergoes surface oxidation in the first several CV scans while simultaneously catalyzing ORR (Figure S9). The anodic peaks corresponding to this oxidative surface reconstruction are observed at the same potential as observed in the absence of $O_2$ (Figure 1d) and upon completion of the surface self-limiting reconstruction, the film displays robust ORR catalysis with an onset potential (defined as the potential corresponding to −0.1 mA cm$^{-2}$ geometric current density) of 0.77 V (Figure 3a). Unlike $Ni_3S_2$, initial CV scans of as-prepared and thermally annealed NiS display no oxidative features and are active for ORR catalysis with an onset potential of 0.78 V, very similar to that of $Ni_3S_2$ upon surface transformation. Steady-state chronoamperograms for $Ni_3S_2$ and as-prepared and annealed NiS were recorded at various applied potentials to construct Tafel plots of overpotential versus the logarithm of the activation-controlled current density (Figure 3b). Tafel plots for $Ni_3S_2$, as-prepared NiS, and annealed NiS samples nearly overall over the entire 1.5 decade range of Tafel data collection, illustrating the similarity in their intrinsic ORR activity. Together, these data indicate that the $Ni_3S_2$ undergoes an oxidative transformation to generate an active amorphous surface phase of nominal NiS stoichiometry.

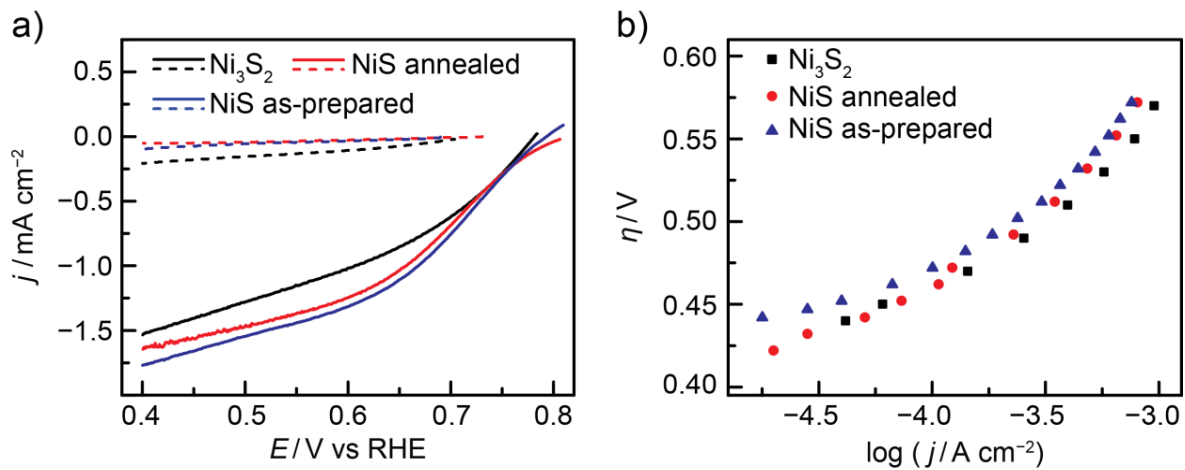

**Figure 3**. a) Linear sweep (5 mV s$^{-1}$ scan rate) voltammograms (LSVs) and b) Tafel plots of $Ni_3S_2$ (black), as-prepared NiS (blue), and annealed NiS (red). LSVs (a, solid lines) and steady state measurements (b) were recorded in $O_2$-saturated 1 M sodium phosphate electrolyte, pH 7, at a rotation rate of 2000 rpm. LSVs recorded $N_2$-saturated electrolyte are plotted in dashed lines.

The high activity of the amorphous surface films raises interest in identifying the nature of the specific active sites responsible for ORR catalysis. Given the rich Ni-S phase diagram, we expect a wide variety of possible active sites, especially in the amorphous surface layer, and thus, it is necessary to develop general principles that connect local structure to electrocatalytic activity. In order to probe the possible active sites, we initially performed density functional theory calculations of ORR activity on low-index facets of several crystalline nickel sulfide phases – $NiS_2$, $Ni_3S_4$, NiS, $Ni_9S_8$, $Ni_3S_2$ (Figure S11). The spectrum of local active site structures on these crystalline nickel sulfide phases and their corresponding calculated electrocatalytic activity was used to formulate a correlation between the local coordination of active sites and their associated ORR activity. This correlation was used to probe the origin of the high activity of the restructured amorphous surface phase.

In the associative mechanism for ORR, the initial reduction of $O_2$ leads to OOH$^*$ and subsequent oxygen-oxygen bond cleavage yields O$^*$ and OH$^*$, which are reduced to water.(Nørskov et al., 2004) The electrocatalytic activity for ORR is thus determined by the free energies of adsorption of OOH$^*$, OH$^*$ and O$^*$. We computed the adsorption free energies of these intermediates on the known crystalline phases of the nickel sulfide family. We observe scaling between the adsorption energies of OOH$^*$ and OH$^*$ on the various Ni-S phases. The scaling has a slope of 1 with an intercept close to 3.04, shown in Fig 4a. The slope of 1 is the same as that observed on metallic surfaces(Abild-Pedersen et al., 2007; Montemore and Medlin, 2014), nanoparticles(Fu et al., 2013) and various transition-metal compounds such as oxides, carbides, nitrides and sulfides(Fernández et al., 2008), and can be rationalized based on bond-order conservation principles(Bell and Shustorovich, 1990; Shustorovich, 1986). It is worth emphasizing that the intercept of the scaling relation differs depending on the degree of solvation around the intermediates.(Viswanathan and Hansen, 2014) On metal surfaces, water stabilizes OH$^*$ much more relative to OOH$^*$; on oxide surfaces, however, there is no additional stabilization for either OOH$^*$ or OH$^*$. As would be expected, the calculated intercept is close to that observed on oxide surfaces.(Calle-Vallejo et al., 2013)

An important consequence of scaling is that the activity is governed by the free energy of one reaction intermediate(Calle-Vallejo and Koper, 2012; Man et al., 2011; Stephens et al., 2012), chosen to be the free energy of OH for this study. From the reaction free energies for the associative mechanism, we can determine the limiting potential, $U_L$, which is the highest potential at which the reaction mechanism is downhill in free energy. This thermodynamic approach has been used successfully in understanding trends in activity(Viswanathan et al., 2012a) and selectivity(Krishnamurthy et al., 2016; Viswanathan et al., 2012b) in oxygen reduction and oxygen evolution(Man et al., 2011). Using scaling, we can plot the limiting potential, $U_L$, as a function of the adsorption energy of OH, $G_{OH^*}$, and is shown in Fig. 5. The strong-binding (left) leg is determined by the removal of OH$^*$ and the weak-binding leg is determined by activation of $O_2$ as $OOH^*$. A recently developed exchange correlation functional, Bayesian Error Estimation Functional with van der Waals correlation (BEEF-vdW)(Wellendorff et al., 2012), allows for estimating uncertainty in the predicted limiting potentials(Deshpande et al., 2016) and a more detailed description of the approach is given in the SI. Using this approach, a quantity known as the expected limiting potential, $U_{EL}$, can be extracted which gives the value of limiting potential taking into account the uncertainty associated with density functional theory calculations.

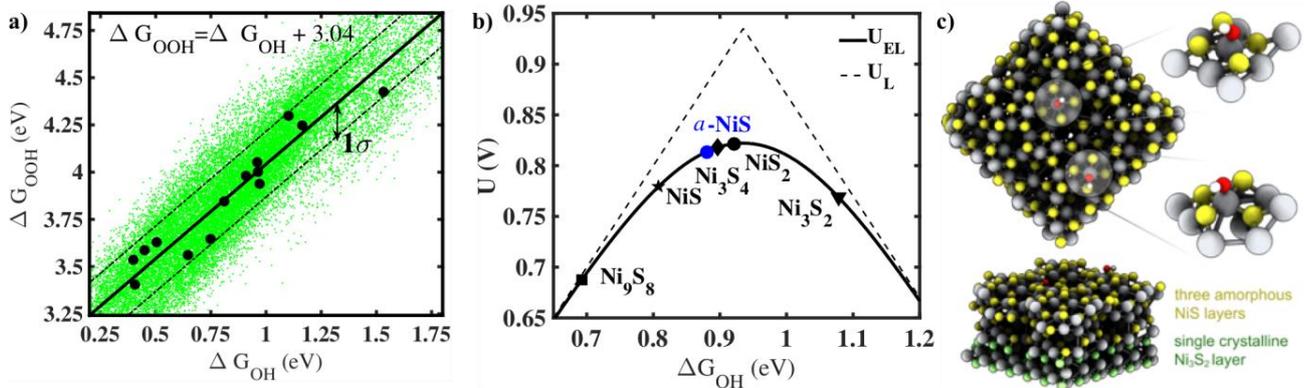

**Figure 4:** a) Scaling between adsorption free energies of OH* and OOH* on the various stable Ni-S phases. The black dots represent the adsorption energies and the green dots represents the ensemble of adsorption free energies obtained from the BEEF-vdW functional that enables error estimation. The best-fit line is given by $\Delta G_{OOH} = \Delta G_{OH} + 3.04$, and the intercept has a standard deviation of 0.17 eV. The black dotted line represents the $1\sigma$ line. b) ORR activity volcano of nickel sulfide phases showing the expected limiting potential (bold line) and the liming potential (dashed line) obtained from a thermodynamic analysis, as a function of the DFT-calculated adsorption free energy of the intermediate OH*. The analysis suggests that multiple Ni-S phases exhibit very high ORR activity ($NiS_2$, NiS, $Ni_3S_4$ and $a$-NiS). $a$-NiS represents the amorphous Ni:S 1:1 phase. c) $a$-NiS structure constructed from ab initio simulated annealing.

Based on DFT calculations, we find that Ni(111) binds $OH^*$ very strongly and lies on the strong binding leg. The addition of sulfur leads to a weakening of $OH^*$ bonding to the active sites, which we find are the Ni centers. The DFT-computed adsorption free energy of $OH^*$ for the various Ni-S phases are shown in Fig 4b and we find that several phases possess surface sites that have high activity. This raises the important question of what common structural features exist among these surface sites. In order to probe these features, we analyzed the structural environment around the active site in search of a structure-energy-activity descriptor. Such descriptors have been identified for metal electrocatalysts for ORR(Calle-Vallejo et al., 2015a, 2015b; Friebel et al., 2012; Jackson et al., 2014)

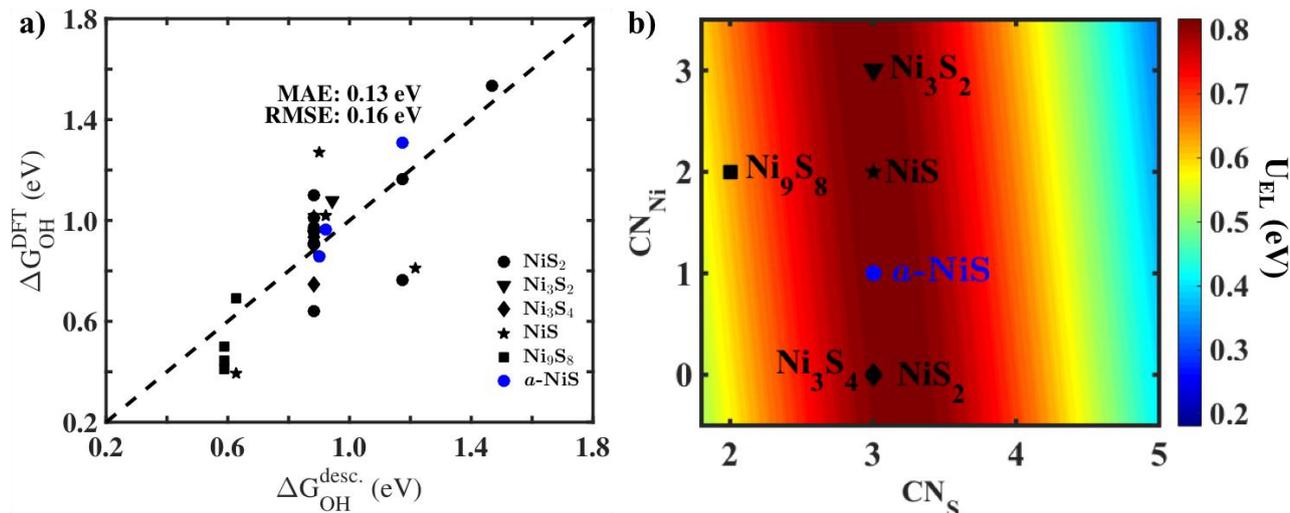

**Figure 5:** a) Correlation between DFT-calculated free energy of adsorption of OH* and that obtained from the structure descriptor, based on the coordination numbers of the nearest neighbor nickel and sulfur atoms. The dashed line represents perfect correlation between the structure descriptor and the DFT-computed $\Delta G_{OH}$. c) Structure-activity contour plot of the expected limiting potential, $U_{EL}$, based on the structure descriptor, $\Delta G_{OH} = 0.29(0.08 CN_{N(Ni)} + CN_{N(S)})$, from the coordination numbers of the nearest neighbor nickel and sulfur atoms. The markers are plotted based on the local coordination around the active site for the various Ni-S phases. It is worth highlighting that all the highly active phases exhibit active sites with three nearest-neighbor sulfur atoms albeit with different number of nickel neighbors.

We characterized the local environment of active sites through the coordination numbers of nearest neighboring nickel and sulfur atoms, $CN_{N(Ni)}$ and $CN_{N(S)}$. We constructed the simplest possible model which linearly depends on these coordination numbers, i.e. $\Delta G_{OH} = \alpha_1 CN_{N(Ni)} + \alpha_2 CN_{N(S)}$. We then carried out a least-squares fit involving all the calculated surface sites. We find, quite remarkably, a structure-energy descriptor relation: $G_{OH} = 0.28(CN_S + 0.08 CN_{Ni})$ which has a mean absolute error (MAE) of 0.13 eV, shown in Fig 5a. This relation shows that the strong covalent Ni-S bonding affects the OH* adsorption much more than the weak Ni-Ni metallic bonding. Therefore, to a first approximation, a simple geometric counting of S neighbors around the Ni surface site provides a measure of the ORR activity.

Using this relation, we can rephrase the challenge of finding active ORR catalysts into finding surface sites that have optimal local coordination. Our analysis suggests that having 3 sulfur atoms bonded to the active site leads to high activity (Figure S18 and S19). Remarkably, all the most active nickel sulfide (Figure 4b) phases possess this local coordination environment (Figure 5b), albeit with different Ni nearest neighbors, and highlights the robustness of the simple structure-energy-activity descriptor identified here. To further validate and probe the local effect, we considered adsorption on an amorphous structure (Figure 4c) with a composition of Ni:S 1.0:1.0, prepared through simulated annealing (see SI for computational details). We find that the same structure-energy descriptor based on local environment determines the activity of sites on the amorphous structure (Table S2) with an average prediction error within the uncertainty of DFT-calculated binding energies.

This analysis suggests that surface sites, with suitable local coordination, can lead to high electrocatalytic activity and that several nickel sulfide phases possess such surface sites. These results imply that oxidative surface transformation of $Ni_3S_2$ serves to generate persistent, active Ni sites bonded to three sulfur atoms, which support high ORR activity. The identification of simple structure-energy-activity descriptors allows for the rapid identification of candidate materials and can accelerate the discovery of Pt-free electrocatalysts for ORR. Given the rich surface dynamics present under electrochemical conditions, the identification of structure-energy-activity descriptors allows for a robust estimate of the activity of the non-equilibrium surface structures that persist under reaction conditions.

**Conclusions**

We have established that Ni$_3$S$_2$ undergoes an oxidative self-limiting surface reconstruction *in situ* under the conditions of ORR catalysis in neutral aqueous electrolytes. This reconstruction serves to generate an approximately 2 nm conformal amorphous surface film with an approximate Ni:S ratio of 1:1. The surface restructured Ni$_3$S$_2$ exhibits similar ORR activity to independently electrodeposited NiS phases, indicating that the reconstructed surface provides the active sites for ORR catalysis. DFT calculations reveal that the OH$^*$ adsorption energy (G$_{OH^*}$) on the *a*-NiS surface phase reflects the distribution in local coordination environments at surface exposed Ni sites. Using an array of crystalline model structures, we have developed, for the first time, a coordination chemistry descriptor for ORR activity on nickel sulfide materials, which highlights the key role of the local coordination environment of the active site. This descriptor can be used to rationalize reactivity trends in amorphous materials and provides a basis for the design of improved first-row transition metal chalcogenides for ORR catalysis.

## Methods

**Bulk Ni$_3$S$_2$ Synthesis.** Phase pure Ni$_3$S$_2$ was prepared by solid-state synthesis.(Tare and Wagner, 1983) 15 mmol Ni powder (0.88 g) and 10 mmol S pieces (0.32 g) were thoroughly mixed and ground in mortar and pestle. The mixture was sealed under vacuum inside a quartz tube and heated for 24 h at 600 ˚C followed by 4 h at 800 ˚C. The resulting product was cooled to room temperature, ground to a fine powder first by mortar and pestle and then ball milled for 1 h. The Ni$_3$S$_2$ electrode was prepared by casting 50 μL of a 10 mg mL$^{-1}$ Ni$_3$S$_2$ suspension in ethanol onto a Au RDE to fully cover the Au surface. After drying in air, the electrode was pressed by a pellet press to obtain a flat surface of Ni$_3$S$_2$.

**Ni$_3$S$_2$ Nanoparticle Synthesis.**(Chi et al., 2012) To a 50 mL three-neck flask, 200 mg nickel (II) 2,4-pentanedionate and 10 mL 1,5-pentanediol was added. Under stirring, a solution of 75.9 mg cystamine dihydrochloride in 10 mL 1,5-pentanediol was added to the nickel solution. The resulting mixture was degassed with N$_2$ and heated under N$_2$ at reflux for 30 min. After cooling to room temperature, the suspended material was isolated via centrifugation at 5000 rpm for 5 min. The crude product was then washed twice using 5 mL 2-propanol and then twice with 5 mL acetone. In each washing step, the sample was re-suspended with a vortex mixer and re-isolated via centrifugation at 5000 rpm for 5 minutes. The product was then dried in air at 60 ˚C for 3 h and stored in the glovebox.

**NiS Film Electrodeposition.**(Jana et al., 2014; Sun et al., 2013) 50 ml of 0.1 M NiCl$_2$·6H$_2$O solution was mixed with 10 ml of 0.1 M aqueous solution of D-(-)-tartaric acid and stirred for 20 minutes. Then, 25 ml of 0.1 M thioacetamide solution was added and stirred for an additional 10 minutes. The final volume of the solution was adjusted to 100 ml with water. The pH of this working solution was maintained at 5.5 by adding 28% ammonium hydroxide solution. A polished Au RDE was used as the substrate and working electrode and together with a Pt mesh counter electrode and an Ag/AgCl reference electrode, an electrochemistry cell was constructed. Consecutive cyclic voltammetry scans from −0.38 to 0.72 V vs RHE were performed at a scan rate of 5 mV s$^{-1}$. During the deposition, the working electrode was rotated at 600 rpm and the solution was continuously bubbled with N$_2$. After 10 cycles, the Au RDE working electrode was removed from the deposition bath, gently washed with copious amount of miliQ water, and dried at room temperature in air. The as-prepared NiS films were annealed at 250 ˚C for 3 h in an N$_2$ atmosphere.

**Physical characterization.** XRD, XPS, SEM, TEM and EDS were conducted to reveal the composition, phase, morphology and structure of each material under investigation (See Supplementary Methods for further details).

**Electrochemical characterizations.** Cyclic voltammetry (CV), chronoamperometry (CA), and potentiostatic Tafel data were collected using a Biologic VSP 16-channel potentiostat or a Gamry Reference 600 potentiostat (See Supplementary Methods for further details).

**Computation.** Density Functional Theory (DFT) calculations were performed using GPAW(Mortensen et al., 2005) with the BEEF-vdW exchange-correlation functional(Wellendorff et al., 2012) using the Atomic Simulation Environment (ASE)(Bahn and Jacobsen, 2002). Core electrons were described using the Projector Augmented Wave Function (PAW), and k-points were represented using Monkhorst Pack grids. All calculations were performed with a grid spacing of 0.18 Å, and converged with a force criterion of < 0.05 eV/ Å. A 6 × 6 × 1 k-point grid was used for a unit cell having 3 atoms each in the $x$ and $y$ directions and with 4 layers in the z direction. The bottom two layers were fixed and the remaining were allowed to be relaxed. The free energies of oxygen intermediates were calculated using DFT, at standard conditions, and at a potential of 0 V versus the Reversible Hydrogen Electrode by incorporating entropy contributions and zero-point energy. The effect of potential U is included by shifting the free energy of an electron by −eU.

**Supplemental Information**
Supplemental Information includes 24 figures and 2 tables.

**Author Contributions**
B.Y. and D.K. are co-first authors. Y.S. and B.Y. conceived of and designed the experimental investigations. B.Y. performed the experiments. V.V., D.K., C.H.H., and S.D. conceived of and designed the computational investigations. D.K., C.H.H., and S.D. performed the computations. Y.S., V.V., B.Y., D.K., and C.H.H analyzed the data and wrote the paper.

**Acknowledgements**
We gratefully acknowledge Dr. Shoji Hall, Anna Wuttig, Youngmin Yoon, and R. Soyoung Kim for helpful discussions. We gratefully acknowledge Dr. Yong Zhang for assistance with HRTEM data collection. The experimental portion of this work was supported by the NSF under award CHE-1454060, and by the MIT Department of Chemistry through junior faculty funds for Y.S. This work made use of the MRSEC Shared Experimental Facilities at MIT, which is supported in part by the NSF under award DMR-0819762. The computational portion of this work was supported by NSF under award CBET- 1554273.

# Surface Restructuring of Nickel Sulfide Generates Optimally-Coordinated Active Sites for ORR Catalysis

Bing Yan, Christopher H. Hendon and Yogesh Surendranath[*]

Department of Chemistry, Massachusetts Institute of Technology, 77 Massachusetts Avenue, Cambridge, Massachusetts 02139-4307, United States

Dilip Krishnamurthy, Siddharth Deshpande, Venkatasubramanian Viswanathan[*]

Department of Mechanical Engineering, Carnegie Mellon University, 5000 Forbes Avenue, Pittsburgh, Pennsylvania 15213, United States

**yogi@mit.edu**

**Experimental Methods:**

**Materials.** Nickel powder, 120 mesh (99.996%, Alfa Aesar), sulfur pieces (99.999%, Alfa Aesar), nickel (II) nitrate hexahydrate (99.9985%, Strem Chemicals), nickel (II) 2,4-pentanedionate (95%, Alfa Aesar), cystamine dihydrochloride (97+%, Alfa Aesar), 1,5-pentanediol (97%, Alfa Aesar), nickel (II) chloride hexahydrate (99.95%, Alfa Aesar), D-(-)-tartaric acid (99%, Alfa Aesar), thioacetamide (98%, Alfa Aesar), ammonium hydroxide solution, 28% in $H_2O$ (≥99.99%, Sigma-Aldrich), sodium phosphate dibasic, anhydrous (≥99.999%, Sigma-Aldrich), sodium phosphate monobasic, anhydrous (≥99.999%, Sigma-Aldrich), perchloric acid (99.999%, Sigma-Aldrich), hydrochloric acid (ACS grade, Ward's Science), phosphoric acid ((≥99.999%, Sigma-Aldrich), and sodium hydroxide (99.99%, Sigma-Aldrich) were used as received without further purification. 1 M sodium phosphate (NaPi) solutions were adjusted to the desired pH with addition of perchloric acid or sodium hydroxide. All electrolyte solutions were prepared with reagent grade water (Millipore Type 1, 18MΩ-cm resistivity).

**General Electrochemical Methods.** Experiments were conducted using a Biologic VSP 16-channel potentiostat or a Gamry Reference 600 potentiostat. Electrochemical measurements were performed in a three-electrode electrochemical cell with a porous glass frit separating the working and auxiliary compartments. The working compartment typically consisted of 30 mL, and the counter compartment consisted of 1.5 mL of electrolyte solution. A high surface area Pt-mesh (Alfa Aesar, 99.997%) was used as the counter electrode. Ag/AgCl reference electrodes stored in saturated NaCl solution were used for electrodeposition of NiS, $Hg/Hg_2SO_4$ reference electrodes stored in saturated $Na_2SO_4$ solution were used for electrolysis under acidic and neutral conditions, and Hg/HgO reference electrodes stored in 1 M NaOH solution were used for electrolysis under alkaline conditions. Unless otherwise noted, potentials are reported in V vs RHE using the following conversions: $E$(RHE) = $E$(Ag/AgCl) + 0.197 V + 0.059(pH), $E$(Hg/Hg$_2$SO$_4$) + 0.656 V + 0.059(pH), or $E$(Hg/HgO) + 0.140 V + 0.059(pH). All experiments were performed at ambient temperature, (21 ± 1 °C). Cyclic voltammetry and steady state data were collected on a rotating disk electrode using a Pine Instruments MSR rotator and a 5 mm diameter Au RDE (Pine).

**X-Ray Diffraction (XRD).** Powder XRD patterns of $Ni_3S_2$ powders prepared by solid-state synthesis and grazing incidence XRD patterns of the electrodeposited NiS films were collected on a Rigaku Smartlab Multipurpose Diffractometer with a Cu source (λ = 1.504 Å). The XRD patterns are shown in Figures S1b, S10a, and S10b.

**Transmission Electron Microscopy (TEM).** $Ni_3S_2$ samples (nanoparticles and ball-milled powders) were drop cast onto Au TEM grids and imaged using a JEOL 2010 FEG Analytical Electron Microscope with an accelerating voltage of 200 kV. The Au grids were attached to Au RDEs with conductive carbon tape and polarized in $O_2$ saturated 1 M NaPi electrolyte, pH 7. The CV or CA-treated samples were rinsed with reagent grade water, dried in a stream of $N_2$ and imaged using the JEOL 2010F instrument. Images were taken under scanning transmission electron microscopy (STEM) mode. In STEM mode, energy-dispersive X-ray spectroscopy (EDS) scans were collected at various positions in the sample extending from the amorphous surface layer to the interior of the particle. The images and EDS results are shown in Figures 2, S2b, S3 and S5.

**Scanning Electron Microscopy (SEM).** SEM images of $Ni_3S_2$ powders prepared by solid-state synthesis were recorded using a Zeiss Ultra Plus Field Emission SEM with an acceleration voltage of 5 keV. The SEM image is shown in Figure S2a.

**X-Ray Photoelectron Spectroscopy (XPS).** XPS spectra were collected using a Physical Electronics PHI Versaprobe II XPS with a monochromated Al (1486.6 eV at 45.6 W) X-ray source and a beam diameter of 200.0 μm. For high-resolution scans, a pass energy of 23.5 eV was used. Depth profiling was performed by collecting sequential XPS scans following progressive removal of surface layers via ionized argon beam sputtering. An $Ar^+$ beam sputter energy of 2 kV was used and each sputter cycle lasted 2 min. XPS spectra are shown in Figures 1a, 1b, S4, S10c and S10d.

**Cyclic Voltammetry (CV).** $Ni_3S_2$ electrodes were prepared as described in the methods section of the main text and rinsed with copious amount of miliQ water. The oxidative surface restructuring was probed in $N_2$-saturated acidic, neutral, and alkaline 1 M $NaP_i$ electrolytes. CV scans (5 mV s$^{-1}$ scan rate) were initiated at the open circuit potential, and 10-20 cycles were recorded consecutively without pause while rotating the electrode at 2000 rpm. ORR catalytic activity was assessed in $O_2$-saturated 1 M $NaP_i$ electrolyte, pH 7, by recording CV scans (5 mV s$^{-1}$ scan rate) initiated at the open circuit potential while rotating the electrode at 2000 rpm. Uncompensated resistances were measured prior to each experiment and typically ranged from 15-20 Ω in the 1 M phosphate solutions, leading to uncompensated Ohmic losses of <5 mV, which were neglected during data processing. All CV scans were recorded without *i*R compensation and produced the data shown in Figures 1d, 3a, S6a, S7, and S9.

**Chronoamperometry (CA).** $Ni_3S_2$ nanoparticles supported on Au TEM grids were polarized at 0.7 V vs RHE in 1 M NaPi electrolyte, pH 7, for 23 s, 2 min, and 5 min to investigate the oxidative surface restructuring. The resulting chronoamperograms are shown in Figure S6c.

**Potentiostatic Tafel Data Collection.** Steady state current-potential data were obtained by conducting controlled potential electrolyses in $O_2$-saturated, 1 M $NaP_i$ electrolyte, pH 7.02, at a variety of applied potentials ranging from 0.42 to 0.58 V vs RHE. Uncompensated resistances were measured prior to each experiment and typically ranged from 15-20 Ω in the 1 M phosphate solutions, leading to uncompensated Ohmic losses of <4 mV, which were neglected during data processing. All data were collected at 2000 rpm to minimize mass transport limitations. In all cases, the measured current reached a steady state within 30 s, and the endpoint current value was plotted in the Tafel plots shown in Figure 3b.

**pH Dependence of Oxidative Surface Restructuring.** We recorded CV scans of freshly prepared $Ni_3S_2$ electrodes in $N_2$-saturated, 1 M $NaP_i$ electrolyte, at pH values of 4.7, 7.0, 8.4, and 9.9 (Figure S7). The anodic peak potential corresponding to oxidative restructuring displays a sub-Nernstian dependence on pH: at pH values of 4.7, 7.0, 8.4, and 9.9 the measured peak potentials were $E_{p,a}$ = 0.39, 0.47, 0.52, and 0.56 V, respectively, indicating that the net surface redox reaction has a proton to electron ratio less than 1. Importantly, the integrated anodic charge passed in the first CV trace declined with increasing pH and was found to be 66.3, 30.5, 3.6, and 1.7 mC for pH 4.7, 7.0, 8.4, and 9.9, respectively. Unlike the behavior observed at neutral or acidic pH, under alkaline conditions, the anodic feature does not decline to a baseline, but rather persists at low level (1.7 mC) even after 20 polarization cycles. The magnitude of this oxidative feature is correlated with the negative switching potential in each scan. As depicted in insets of Figure S7c and S7d, the oxidative peak in question is not observed if the previous negative-going scan is switched at 0.3 V (red solid lines). However, the anodic peak reappears if the negative-going scan is switched at 0.0 V (blue dotted line). Following CV cycling in alkaline electrolyte, electrodes were rinsed with reagent grade water and subsequently cycled in pH 4.7 electrolyte. These CVs scans again display a large anodic peak at 0.39 V which is in line with the original peak potential observed for a freshly prepared $Ni_3S_2$ electrode at pH 4.7.

Together, these observations lead us to postulate that the oxidative surface transformation of $Ni_3S_2$ occur via two concurrent pathways. In acidic electrolytes, the oxidative wave corresponds to electrodissolution of $Ni^{2+}$ via the following equilibrium:

$$Ni_3S_2 \leftrightharpoons Ni^{2+} + 2NiS + 2e^-$$

In contrast, at more alkaline pH values, the generated $Ni^{2+}$ precipitates as $Ni(OH)_2$ on the electrode via the following equilibrium:

$$Ni_3S_2 + 2OH^- \leftrightharpoons Ni(OH)_2 + 2NiS + 2e^-$$

As this second equilibrium does not lead to corrosion of the surface, the $Ni(OH)_2$ and NiS materials generated during the oxidative scan can be back-reduced on the negative-going scan to regenerate $Ni_3S_2$, thereby leading to persistent oxidative feature.[1] Under acidic and neutral conditions, however, the $Ni^{2+}$ diffuses into the bulk electrolyte and preventing back reduction to reform $Ni_3S_2$. As a result, the oxidation peak declines to the background level at pH 4.7 and 7.0, while remaining persistent at pH 8.4 and 9.9.

**Construction of Pourbaix diagram.** Established thermodynamic data[2] for nickel and nickel compounds are listed in Table S1 below and were used to construct a Pourbaix diagram of the Ni-O-S system (Figure S8).

Table S1. Thermodynamic data for Ni compounds

| Materials | $\Delta H$ / kJ mol$^{-1}$ | $\Delta S$ / J K$^{-1}$ mol$^{-1}$ | $\Delta G$ / kJ mol$^{-1}$ |
|---|---|---|---|
| $Ni_3S_2$ heazlewoodite | -217.2±1.6 | 133.5±0.7 | -211.2±1.6 |
| NiS millerite | -94.0±1.0 | 52.97±0.33 | -91.3±1.0 |
| $Ni^{2+}$ | | | -45.77±0.77 |
| NiO | -239.7±0.4 | 38.4±0.4 | -211.66±0.42 |
| $Ni(OH)_2$ beta | -542.3±1.5 | 80.0±0.8 | -457.1±1.4 |

$$E(Ni^{2+}/Ni) = -(0.2372 \pm 0.0040) V$$

$$Ni(OH)_2 + 2H^+ \rightarrow Ni^{2+} + 2H_2O \quad \log_{10} K_s(Ni(OH)_2) = 10.5 \pm 1.3$$

$$H_2S \rightarrow H^+ + HS^- \qquad (1)$$

$$HS^- \rightarrow H^+ + S^{2-} \tag{2}$$

$$pKa_1 = 7.0; pKa_2 = 19.0$$

The solubility data for NiS varies among literature sources,[2] and average value of $\log_{10}K_{sp} = -30$ is used here.

$$Ni_3S_2 \rightarrow Ni^{2+} + H_2S + 2e \tag{a}$$

$$E = -0.088 + \frac{0.059}{2}\log[Ni^{2+}] \quad \text{at pH} < 3$$

$$Ni_3S_2 + 4H^+ \rightarrow Ni^{2+} + NiS + 2e \tag{b}$$

$$E = -0.089 + \frac{0.059}{2}\log[Ni^{2+}] \text{ at } 3 < pH < 5.25 \text{ when } [Ni^{2+}] = 1 \text{ M}$$

$$Ni_3S_2 + 2OH^- \rightarrow Ni(OH)_2 + 2NiS + 2e \tag{c}$$

$$E = 0.221 - 0.059pH \text{ at pH} > 5.25 \text{ when } [Ni^{2+}] = 1 \text{ M}$$

$$Ni \rightarrow Ni^{2+} + 2e \tag{d}$$

$$E = -0.237 + \frac{0.059}{2}\log[Ni^{2+}] \text{ at pH} < 5.25 \text{ when } [Ni^{2+}] = 1 \text{ M}$$

$$Ni^{2+} + 2OH^- \rightarrow Ni(OH)_2 \tag{e}$$

$$pH = 5.25 + \frac{0.059}{2}\log[Ni^{2+}]$$

$$Ni + 2OH^- \rightarrow Ni(OH)_2 + 2e \tag{f}$$

$$E = -0.237 - 0.059pH \text{ at pH} > 5.25 \text{ when } [Ni^{2+}] = 1 \text{ M}$$

The oxidation products of $Ni_3S_2$ depend on the acidity of the solution: under strongly acidic conditions, $Ni^{2+}$ and $H_2S$ are generated; from mildly acidic to neutral pH, $Ni^{2+}$ and NiS result; from neutral to alkaline conditions, $Ni(OH)_2$ and NiS are the products. The thermodynamic potentials of the reaction depend on $Ni^{2+}$ concentration and pH of the electrolyte. For example, at pH 7 and $[Ni^{2+}] = 10^{-5}$ mol L$^{-1}$, the thermodynamic potential is 0.22 V vs RHE and the product is mainly $Ni^{2+}$ and NiS, consistent with our experimental observations. With the results together, we propose a pathway of the $Ni_3S_2$ surface self-limiting oxidation in neutral media: 1) the $Ni_3S_2$ surface is oxidized upon electrochemical polarization and the product NiS gradually forms a layer on the surface of each crystallite; 2) concurrently, a significant fraction of $Ni^{2+}$ ions oxidatively leach into the solution due to the low Faradaic efficiency of NiS formation; 3) a small amount of $Ni^{2+}$ precipitates as $Ni(OH)_2$ onto the electrode surface. These processes continue until the surface has been completely protected from further corrosion by a conformal amorphous film of NiS and minority $Ni(OH)_2$ species.

**Quantification of Uncertainty in Adsorption Energies.** The recently developed exchange correlation functional, BEEF-vdW (Bayesian Error Estimation Functional)[3], enables uncertainty quantification through its unique error estimation capability built into it. The method uses different types of data sets as the optimal data and fits the GGA exchange enhancement factor, $F_x(s)$ to it.

The expression for the exchange correlation energy for the BEEF-vdW functional is given by

$$E_{xc} = \sum_{m=0}^{M_x-1} a_m E_m^{GGA-x} + \alpha_c E^{LDA-c} + (1-\alpha_c) E^{PBE-c} + E^{nl-c}$$

where $M_x$ represents the degree of the polynomial. The coefficients $a_m$ and $\alpha_c$ are the fitting parameters which are optimized over the data sets. $E_m^{GGA-x}$ represents the GGA exchange energy, $E^{PBE-c}$ and $E^{LDA-c}$ represent the PBE and LDA correlation energies and $E^{nl-c}$ represents the non-local correlation energy obtained from the functional vdW-DF2[4].

BEEF-vdW uses an ensemble of functionals to give an ensemble of DFT energies.[5] After calculating the optimum value for the coefficients $a_m$ and $\alpha_c$, each coefficient is perturbed around its optimal value. From the ensemble of coefficients an ensemble of energies is generated. The ensemble of energies provides a systematic way to calculate the uncertainty associated within the GGA class of functionals for a given calculation. This approach provides a computationally tractable way to estimate the uncertainty associated with a given calculation.

Using this approach, an ensemble of exchange correlation functionals results in an ensemble of adsorption energies, from which the error in the adsorption energy is obtained as the sensitivity of DFT results to the choice of the exchange-correlation functional. Using the BEEF-vdW functional, ensembles of adsorption energies for various intermediates involved in the ORR were calculated. Referencing the adsorption energy of any intermediate with respect to gas phase molecules (here $O_2(g)$) leads to large estimated error.

However, when the reference is changed from the gas phase molecules to a reference phase, for e.g., $NiS_2$ (111), it is found that this leads to much smaller error estimates due to similarity in surface-adsorbate bonding characteristics. This is in agreement with the notion that trends in DFT calculations are more accurate than the individual values due to cancellation of systematic errors. We use the following methodology to get a combined error estimate for the adsorption energies of various intermediates. We illustrate the approach using the example of the intermediate OH*.

First, the ensemble of OH* adsorption energies for a given facet 'X' with respect to the reference system, chosen as $NiS_2$, is calculated, given by

$$\Delta G_{OH}(X|NiS_2(111)) = \Delta G_{OH}(X) - \Delta G_{OH}(NiS_2(111))$$

The distribution for the calculated ensemble of adsorption energies is centered around the mean:

$$\overline{\Delta G_{OH}(X|NiS_2(111))} = \Delta G_{OH}(X|NiS_2(111)) - <\Delta G_{OH}(X|NiS_2(111))>$$

We carry out for all the facets considered in this study and a combined distribution is constructed/:

$$\Delta G_{OH} = \sum_X \overline{\Delta G_{OH}(X|NiS_2(111))}$$

The standard deviation for the combined distribution ($\sigma_{OH}$) is 0.09 eV.

Based on this analysis, we approximate the uncertainty involved in determining the adsorption energy of an intermediate by the standard deviation of the calculated adsorption energy ensemble on the various nickel sulfide facets.

**Construction of Scaling Relationships.** It has been shown that there exists a linear scaling between the adsorption energies of the hydrogenated species, $AH_x$ and the atom A[6]. Using simple bond counting principles it was shown that the slope of the linear scaling is only dependent on the valency and not on the specific facet considered.[7] This has been shown for metallic surfaces, nanoparticles and various transition-metal compounds such as oxides, carbides, nitrides and sulfides.

We have explored trends in adsorption energy between OH* and OOH*, and observe that the adsorption free energies between OH* and OOH* scale with each other independent of the facet of all nickel sulfide phases considered. We explore the uncertainty in the intercept of the scaling relationship. As mentioned in the main text, we assume that the slope between the adsorption energies of the intermediates is one. This arises due to arguments based on the bond order conservation principle. Figure 4a shows the best fit line. The expression for the standard deviation of the scaling intercept used is derived in the following way:

For two random variables, $X$ and $Y$,

$$(\sigma_{X-Y})^2 = E[(X-Y)^2] - (E[(X-Y)])^2$$
$$(\sigma_{X-Y})^2 = E[X^2 + Y^2 - 2XY] - (E[X] - E[Y])^2$$
$$(\sigma_{X-Y})^2 = E[X^2] + E[Y^2] - 2E[XY] - (E[X])^2 - (E[Y])^2 + 2E[X]E[Y]$$
$$(\sigma_{X-Y})^2 = (\sigma_X)^2 + (\sigma_Y)^2 - 2(E[XY] - E[X]E[Y])$$
$$(\sigma_{X-Y})^2 = (\sigma_X)^2 + (\sigma_Y)^2 - 2(\mu_{XY} - \mu_X \mu_Y)$$

The relation, $(\sigma_{OOH^*-OH^*})^2 = (\sigma_{OOH^*})^2 + (\sigma_{OH^*})^2 - 2(\mu_{OOH^*OH^*} - \mu_{OOH^*}\mu_{OH^*})$, was used to generate the standard deviation of the intercept, where $\sigma_{OH^*}$, $\mu_{OH^*}$ and $\sigma_{OOH^*}$, $\mu_{OOH^*}$ are the standard deviations and mean values for $\Delta G_{OH^*}$ and $\Delta G_{OOH^*}$ respectively. $\mu_{OH^*OOH^*}$ is the mean associated with the distribution of $\Delta G_{OH^*} \times \Delta G_{OOH^*}$. $\sigma_{OOH^*-OH^*}$ is the standard deviation for the scaling relationship.

**Determination of the Expected Limiting Potential.** The uncertainty is predicted using a parameter defined[8] as the expected limiting potential, $U_{EL}$, which is the expected value of the limiting potential, $U_L$. The deviation of the expected limiting potential, $U_{EL}$, from the thermodynamic limiting potential, $U_L$ is a qualitative estimate of the prediction uncertainty and can be used to identify trends in predictability. We notice that the expected limiting potential, $U_{EL}$, and the limiting potential, $U_L$ deviate from each other close to the top of the activity volcano, which implies that the activity predictions from the thermodynamic activity volcano become less reliable in this region. Figure 4b provides a way to understand the uncertainties associated with the limiting potentials, however, it does not provide a visual representation of the probability distribution associated with the limiting potential, $U_L$.

To determine the probability distribution, we first determine the probability distribution for the limiting potential $U_L$ as a function of the free energy of the intermediate OH* relative to NiS$_2$(111). Then, we consider a random variable $\Delta G_{OH^*} \sim N(<\Delta G_{OH^*}>, \sigma_{OH^*})$, where, $<\Delta G_{OH^*}>$ is the mean and $\sigma_{OH^*}$ is the standard deviation corresponding to the adsorption energy of the intermediate, OH*. A given value of the mean, $<\Delta G_{OH^*}>$ represents a calculated value of the free energy while the random variable, $\Delta G_{OH^*}$, accounts for the uncertainty of the calculated value. This gives rise to a probability distribution of the limiting potential, $U_L$ as a function of the mean, $<\Delta G_{OH^*}>$. From the probability distribution, the expected limiting potential ($U_{EL}$) is derived, which is the expectation value of the limiting potential, $U_L$. For a given value of $<\Delta G_{OH^*}>$, the expected limiting potential represents the value that would be expected given a large number of experiments on materials with the same calculated value.

**Generation of Amorphous Surface Phases.** To generate a plausible representation of our amorphous experimental surface, we generated a slab model that featured sub-surface crystalline Ni$_3$S$_2$ with ~12 Å of amorphous NiS on the surface (Figure 4c). The amorphous material was generated using an *ab initio* simulated annealing method: beginning with a 2 x 2 expansion of a (100) Ni$_3$S$_2$ slab, we manually removed one third of the Ni atoms from the top three surface layers and subjected the overall structure of simulated annealing, followed by quenching to generate a representative local minimum structure that reflects the inherent heterogeneity of surface exposed Ni sites.

**Table S2.** Adsorption free energies calculated from DFT, and through the structure-energy descriptor, on the three stable surface sites on the simulated amorphous structure with a composition of Ni:S 1.0:1.0.

| Local coordination | | Adsorption Free Energy (eV) | |
|---|---|---|---|
| $CN_{Ni}$ | $CN_S$ | $\Delta G_{OH}^{desc.}$ | $\Delta G_{OH}^{DFT.}$ |
| 1 | 3 | 0.89 | 0.86 |
| 2 | 3 | 0.91 | 0.96 |
| 0 | 4 | 1.16 | 1.31 |

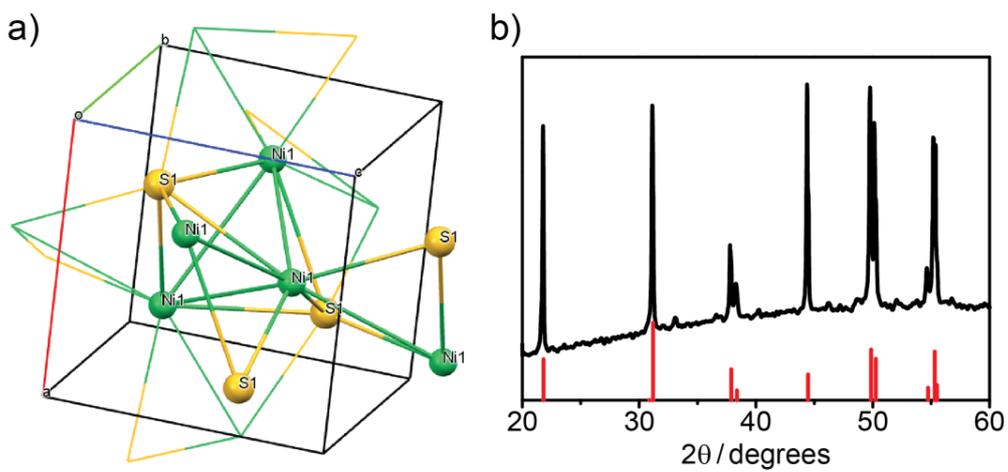

**Figure S1.** a) Crystal structure and b) PXRD pattern of $Ni_3S_2$.[9]

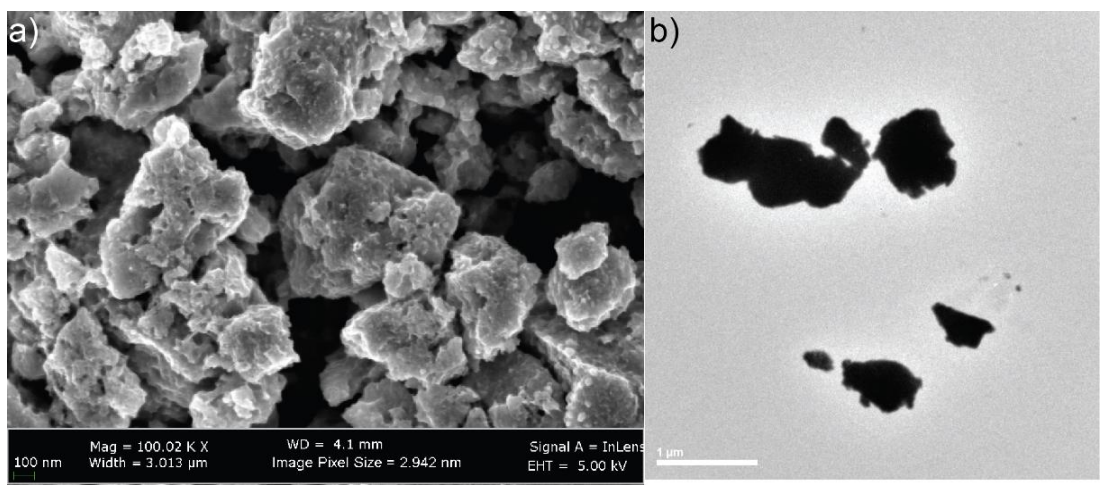

**Figure S2.** a) SEM and b) TEM images of $Ni_3S_2$ particles prepared by solid-state synthesis.

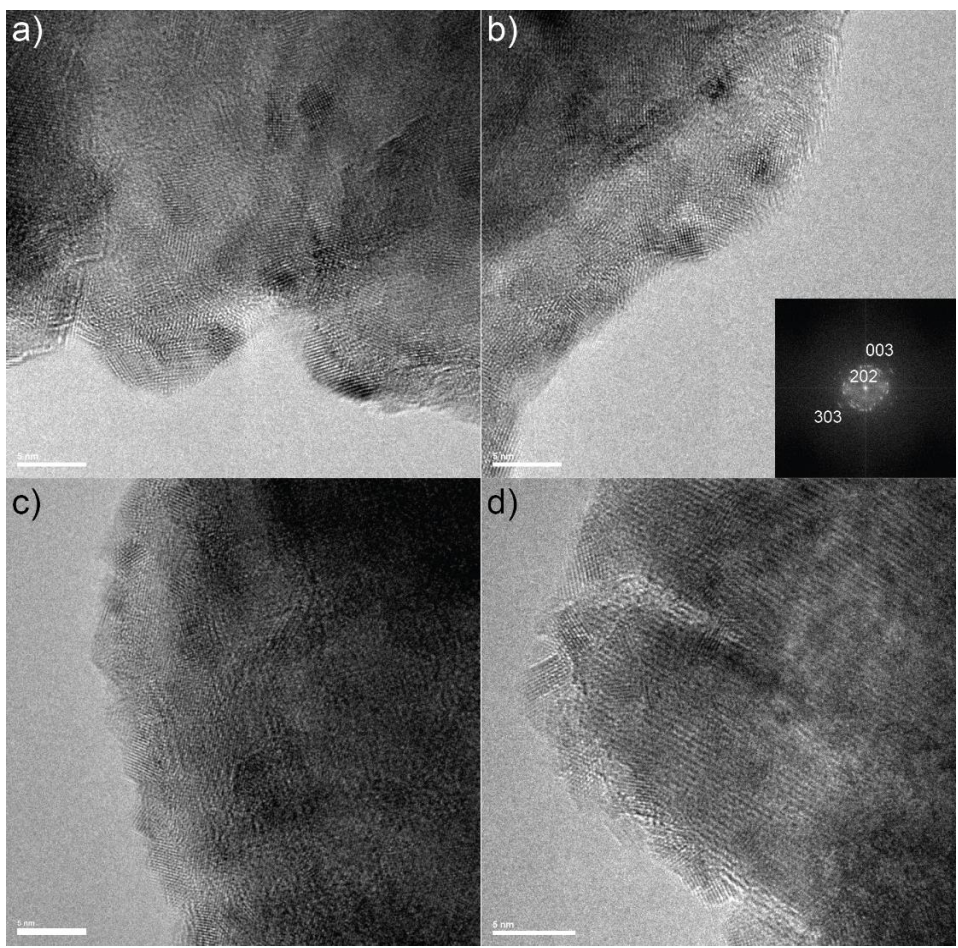

**Figure S3.** Representative TEM images of the as-prepared $Ni_3S_2$ obtained from solid state synthesis (a and b) and nanoparticle synthesis (c and d). The lattice fringes extend to the edge of each crystallite. The inset of panel b) shows the FFT of the image, indicating the polycrystalline nature of the particulates.

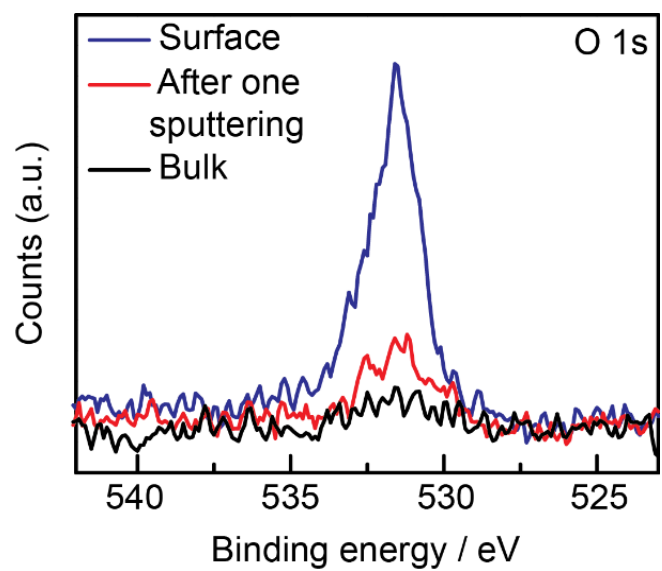

**Figure S4.** XPS of O 1s on the surface (blue) of $Ni_3S_2$ after performing ORR. The peak decreased significantly after one cycle of $Ar^+$ sputtering (red) and finally disappears in the bulk of the sample (black).

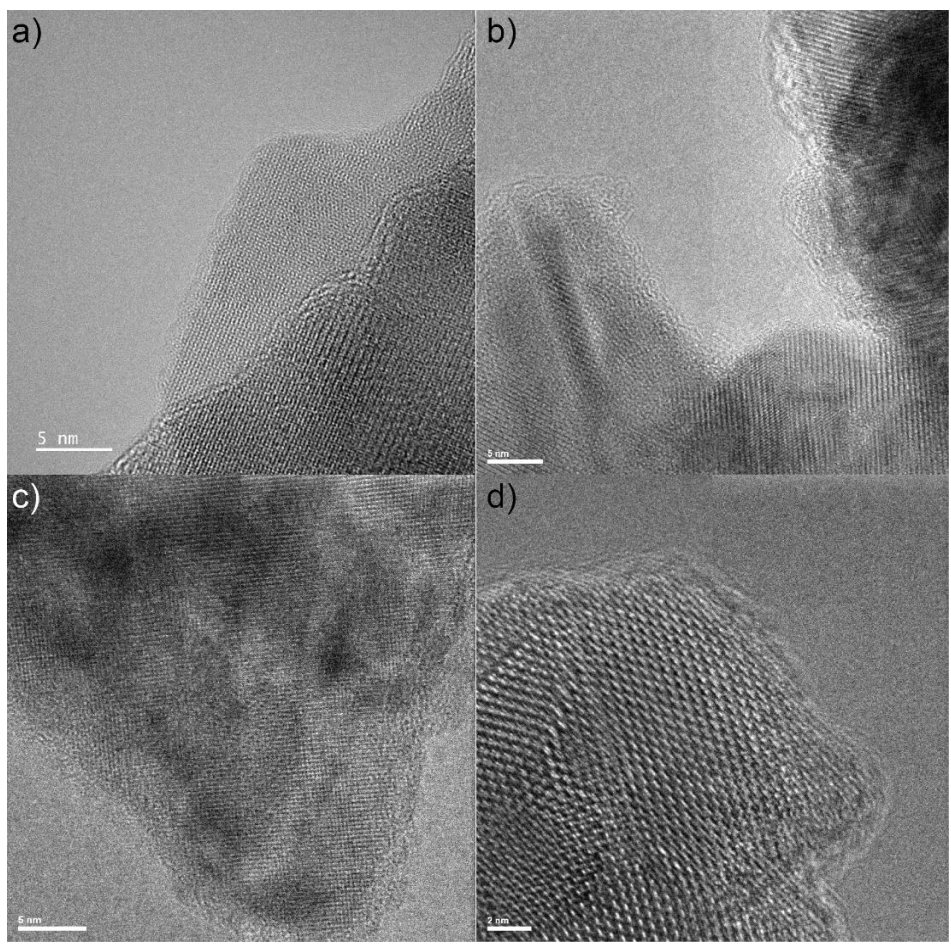

**Figure S5.** Representative high-resolution TEM images on Ni$_3$S$_2$ nanoparticles following surface oxidative transformation. A thin amorphous layer surrounds each nanoparticle.

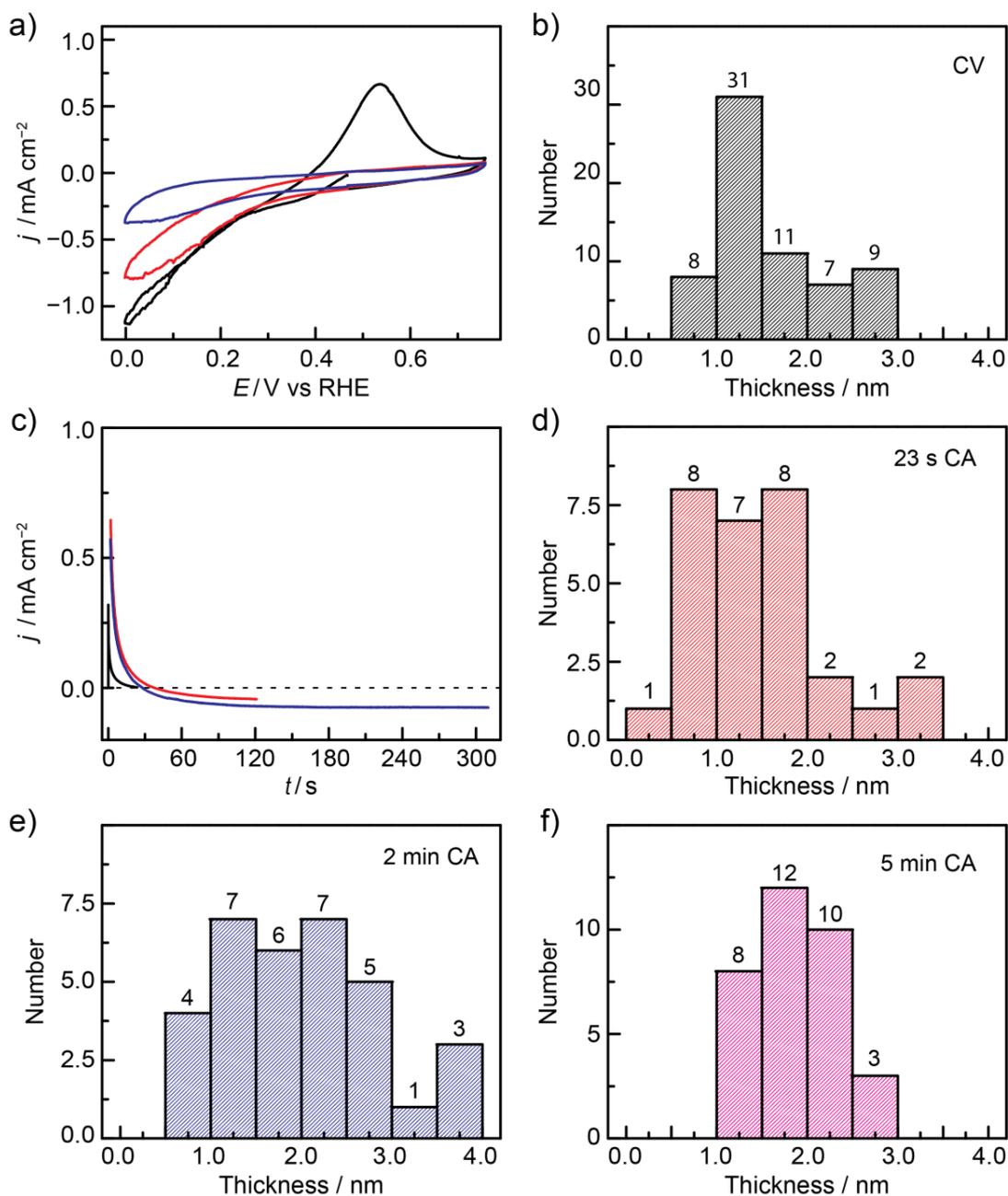

**Figure S6.** Polarization of the Ni$_3$S$_2$ nanoparticles and histograms of the amorphous surface layer thickness. Polarization of Ni$_3$S$_2$ nanoparticles drop cast onto TEM grids by a) cyclic voltammetry (CV) and c) chronoamperometry (CA). In CA, the polarization time was increased from 23 s (b, black), 2 min (b, red) and 5 min (b, blue). The cathodic current density decreases in the CV because particles loosely affixed to the grid detach from the surface over time. The electrochemistry was performed in O$_2$-saturated 1 M NaPi, pH 7, electrolyte. The histograms (b, d-f) were constructed by measuring the amorphous layer thickness across 10-20 particles for each sample.

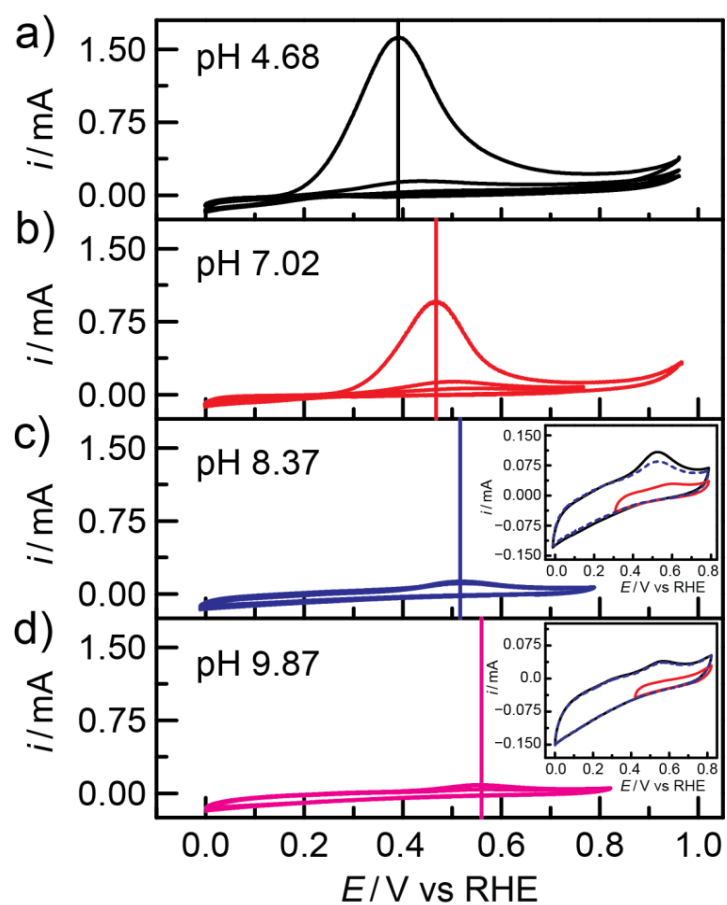

**Figure S7.** pH dependence of the oxidative conversion wave. The slope of −0.03 V pH$^{-1}$ indicates that the oxidation is a sub-Nernstian process.

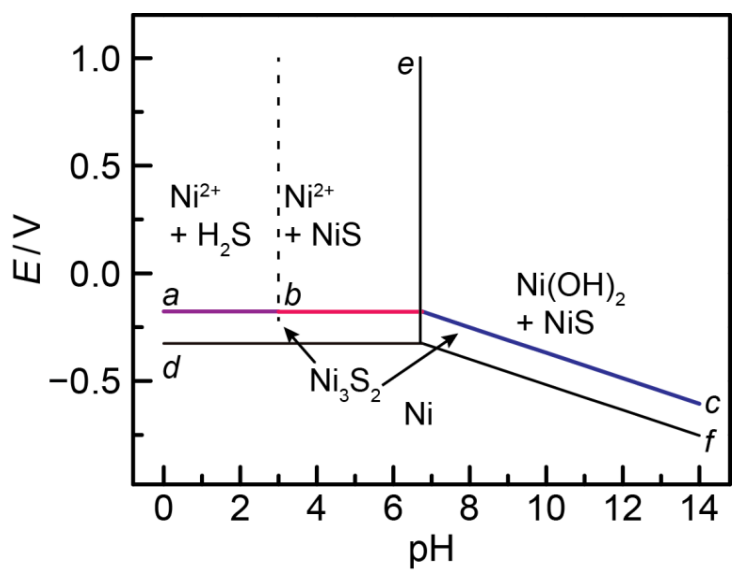

**Figure S8**. Calculated Ni-O-S Pourbaix diagram assuming an equilibrium [$Ni^{2+}$] concentration of 0.001 M. $Ni_3S_2$ is oxidized to $Ni^{2+}$, NiS and/or $Ni(OH)_2$ depending on the acidity of the solution. The letters correspond to the equilibrium outline above.

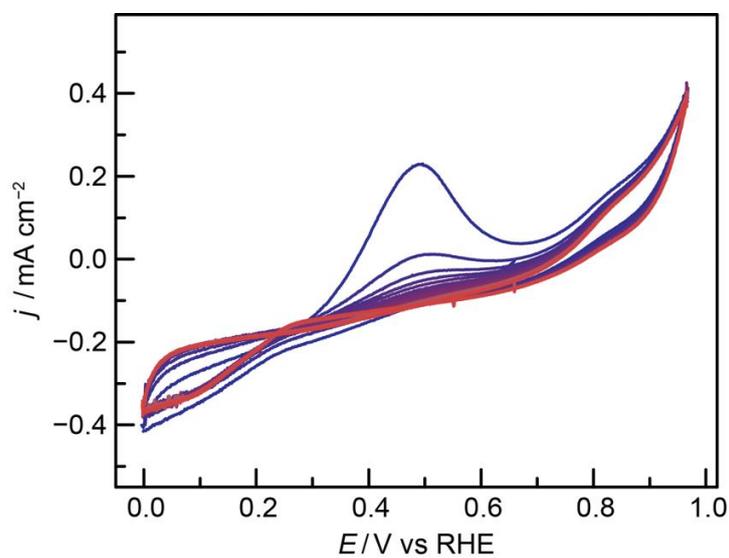

**Figure S9.** CV on a freshly-prepared $Ni_3S_2$ electrode in the presence of $O_2$. The $Ni_3S_2$ electrode undergoes oxidative surface reconstruction while simultaneously catalyzing ORR.

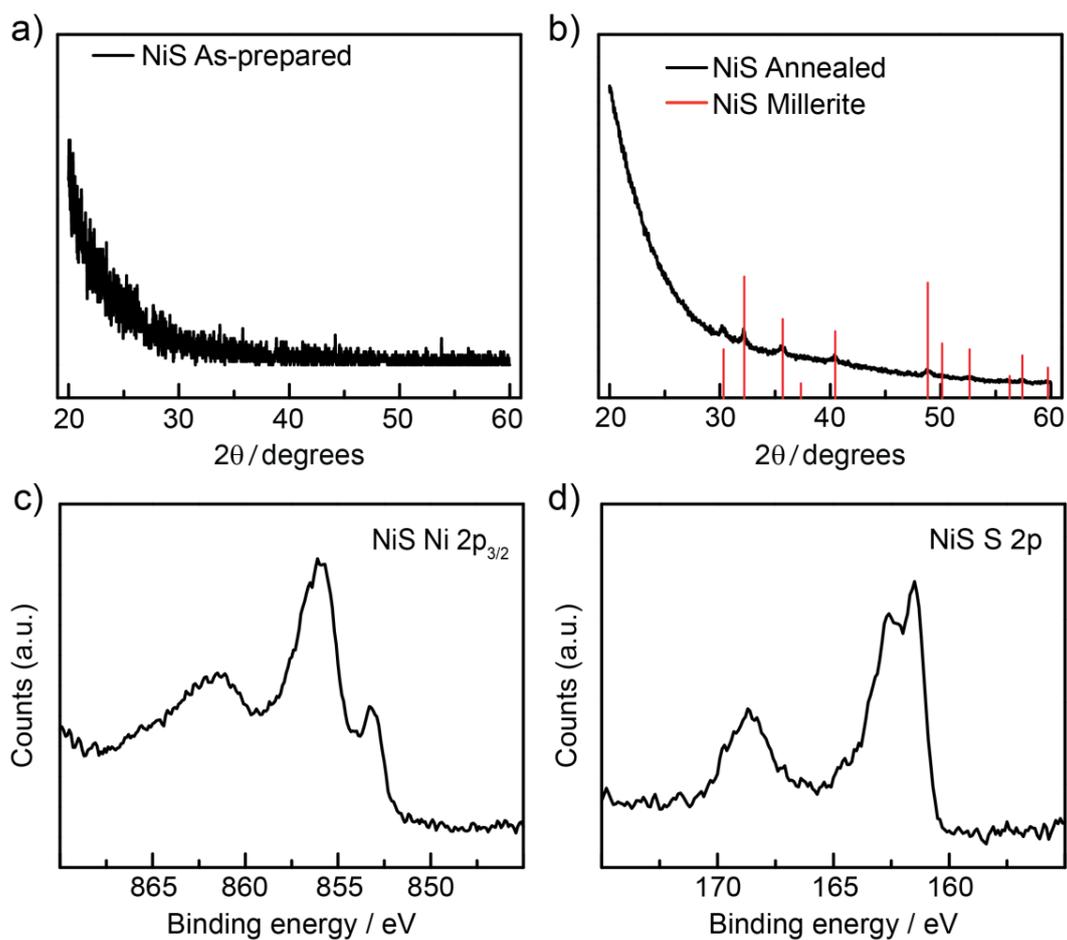

**Figure S10.** GIXD of a) as-prepared NiS and b) annealed NiS. The as-prepared sample does not display crystalline peaks, while the annealed NiS displays XRD peaks consistent with the millerite NiS phase. c) and d) XPS of electrodeposited NiS. The Ni $2p_{3/2}$ peaks are split into two chemical environments: NiO at 856.1 eV and NiS at 853.3 eV. NiO is known to be catalytically inert for ORR.[10]

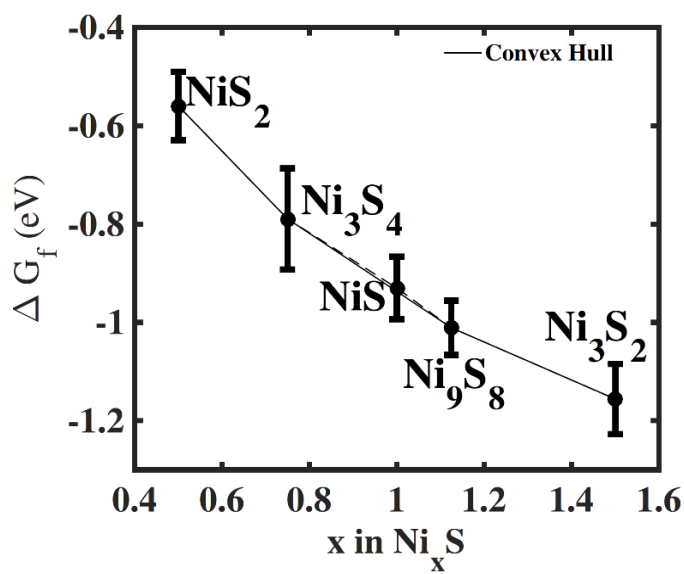

**Figure S11.** Formation free energies calculated using DFT for the five stable phases of nickel sulfides. The error bars represent the standard deviation in the formation free energy calculated using the BEEF-vdW functional.

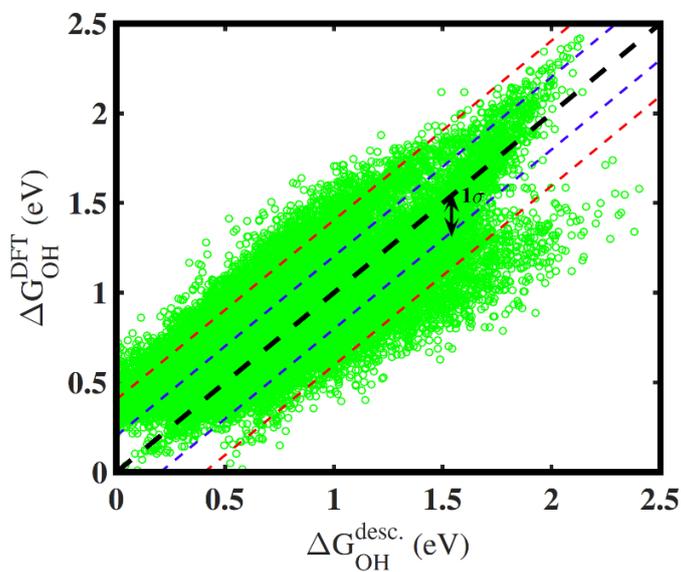

**Figure S12.** Distribution of the correlation between DFT-calculated free energy of adsorption of OH* and that obtained from the structure descriptor, based on the coordination numbers of the nearest neighbor nickel and sulfur atoms. The dashed line represents perfect correlation between the structure descriptor and the DFT-computed $\Delta G_{OH}$. For each of the 2000 functionals within the BEEF-vdW exchange correlation functional, $\Delta G_{OH}^{desc.}$ and $\Delta G_{OH}^{DFT}$ on various Ni-S phases are calculated using the structure descriptor and from density functional theory calculations respectively.

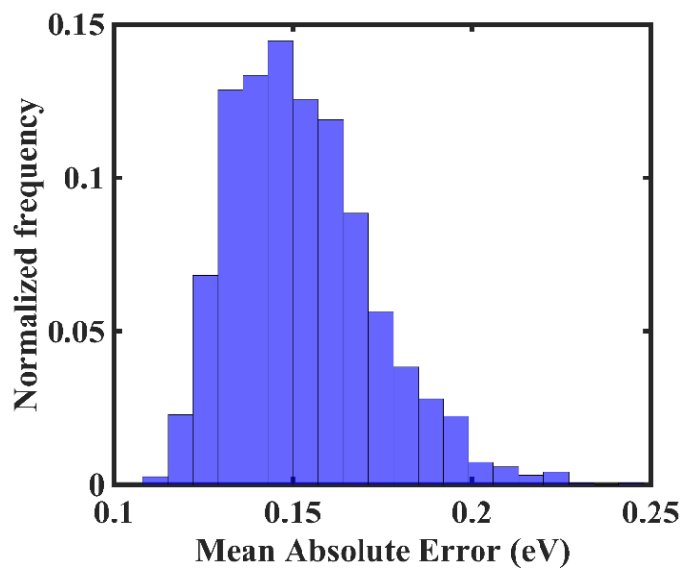

**Figure S13.** Distribution of the Mean Absolute Error ( average $|\Delta G_{OH}^{DFT} - \Delta G_{OH}^{desc.}|$) for the 2000 functionals of the BEEF-vdW exchange correlation functional. $\Delta G_{OH}^{desc.}$ and $\Delta G_{OH}^{DFT}$ on various Ni-S phases are calculated using the structure descriptor and from density functional theory calculations respectively from the corresponding functional.

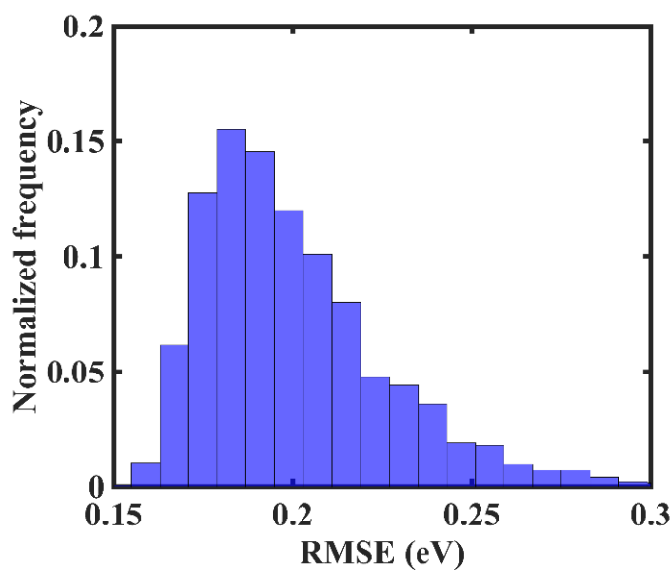

**Figure S14.** Distribution of the Room Mean Square Error (RMSE between $\Delta G_{OH}^{DFT}$ from $\Delta G_{OH}^{desc.}$) for the 2000 functionals of the BEEF-vdW exchange correlation functional. $\Delta G_{OH}^{desc.}$ and $\Delta G_{OH}^{DFT}$ on various Ni-S phases are calculated using the structure descriptor and from density functional theory calculations respectively from the corresponding functional.

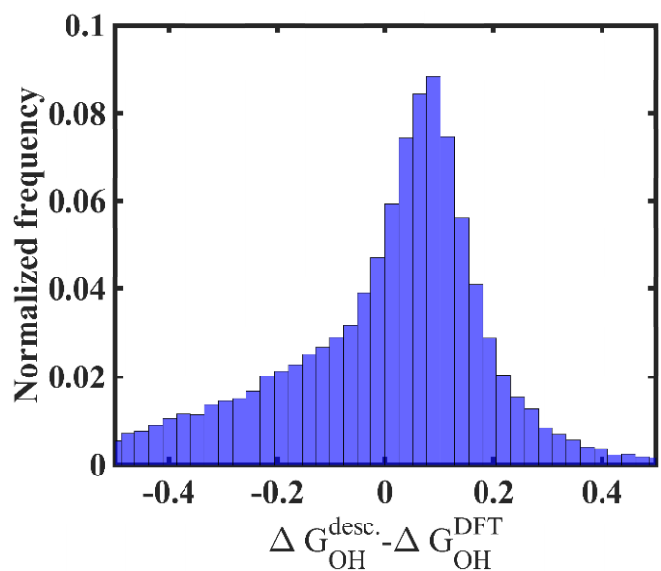

**Figure S15.** Distribution of the difference between $\Delta G_{OH}^{desc}$ and $\Delta G_{OH}^{DFT}$ for various Ni-S phases from all the 2000 functionals within the BEEF-vdW exchange correlation functional. $\Delta G_{OH}^{desc.}$ and $\Delta G_{OH}^{DFT}$ on various Ni-S phases are calculated using the structure descriptor and from density functional theory calculations respectively.

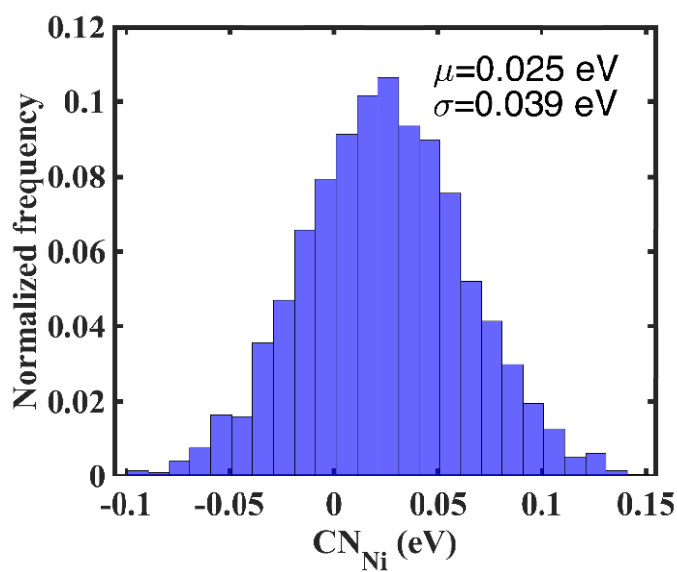

**Figure S16.** Distribution of the adsorption energy contribution per Ni in the nearest shell obtained using the BEEF-vdW functional. Each member of the distribution is obtained by constructing a best-fit structure descriptor based on the nearest shell coordination numbers of Ni and S for one set (out of the ensemble of 2000 functionals) of OH* adsorption energies for all the active sites in various Ni-S phases.

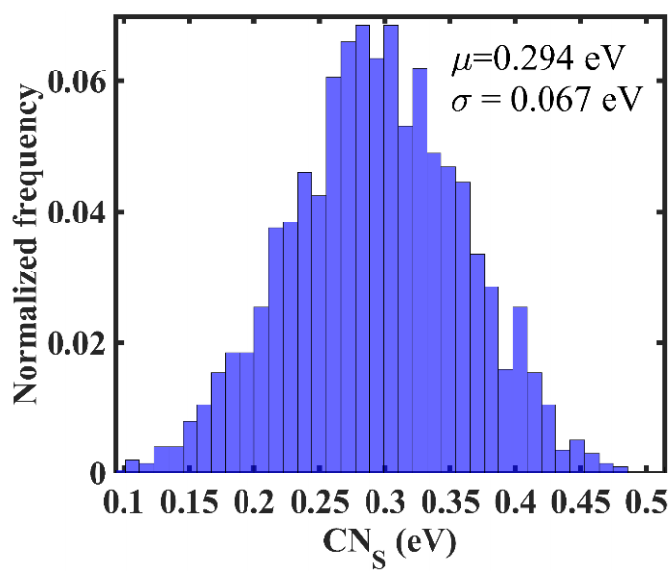

**Figure S17.** Distribution of the adsorption energy contribution per S in the nearest shell obtained using the BEEF-vdW functional. Each member of the distribution is obtained by constructing a best-fit structure descriptor based on the nearest shell coordination numbers of Ni and S for one set (out of the ensemble of 2000 functionals) of OH* adsorption energies for all the active sites in various Ni-S phases.

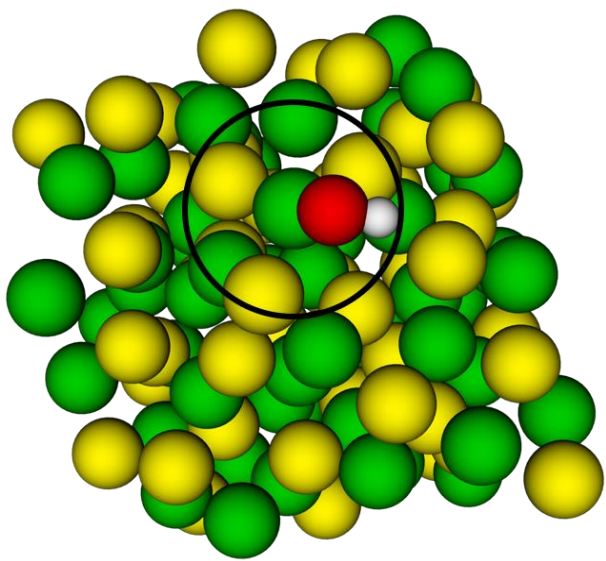

**Figure S18.** The local coordination for the most active site on the simulated amorphous 1:1 Ni:S surface. We observe that it has 3 sulfur atoms in the nearest neighbor shell.

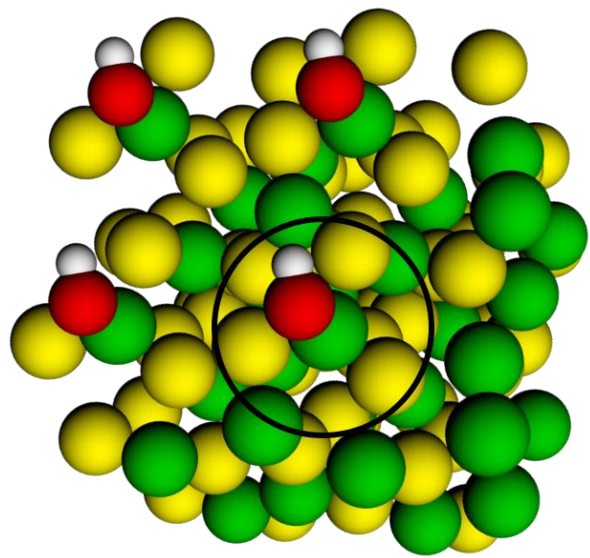

**Figure S19.** The local coordination for the most active site of the NiS$_2$ phase. We notice again that it has 3 sulfur atoms in the nearest neighbor shell.

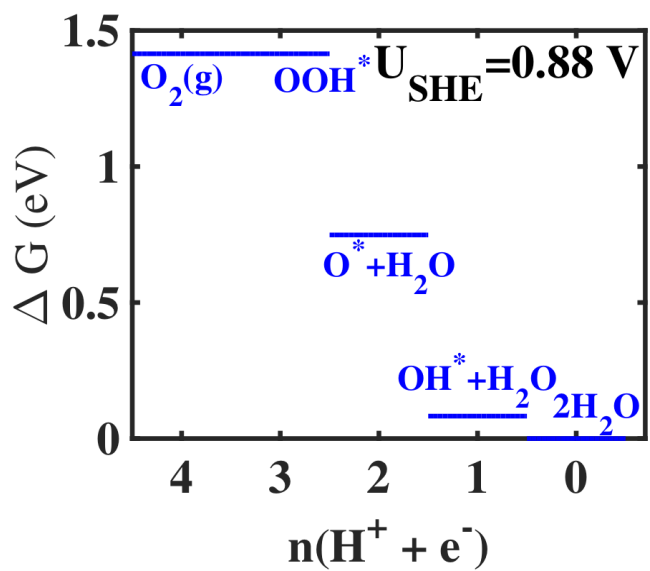

**Figure S20.** Free energy diagram plotted at the limiting potential on the NiS$_2$ phase. The step involving the activation of $O_2$ as OOH* is the potential determining step and the free energy levels of these two intermediates line up at the limiting potential of 0.88 V.

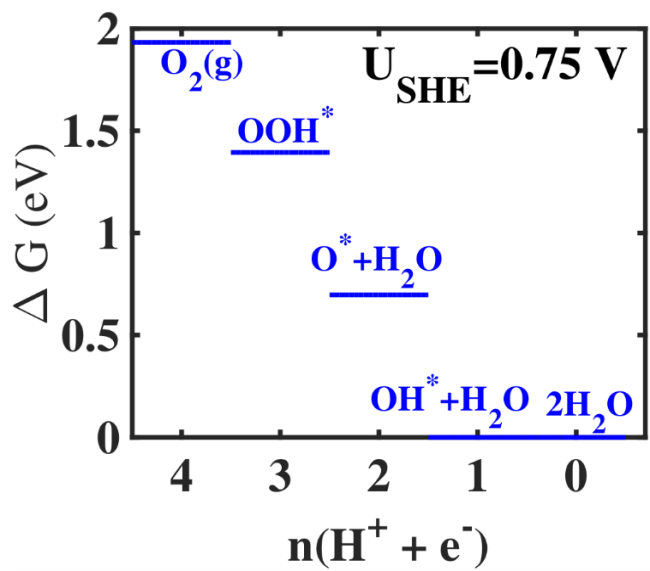

**Figure S21.** Free energy diagram plotted at the limiting potential on the Ni$_3$S$_4$ phase. The step involving the reduction of OH* to H$_2$O is the potential determining step and the free energy levels of these two intermediates line up at the limiting potential of 0.75 V.

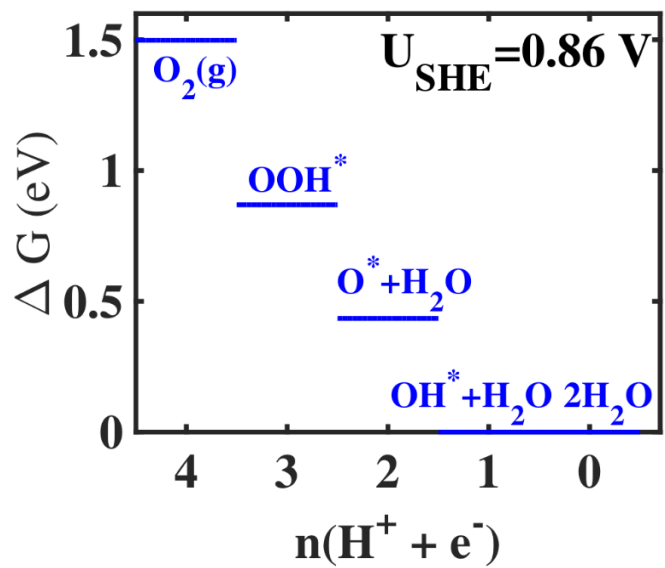

**Figure S22.** Free energy diagram plotted at the limiting potential on the NiS phase. The step involving the reduction of OH* to H₂O is the potential determining step and the free energy levels of these two intermediates line up at the limiting potential of 0.86 V.

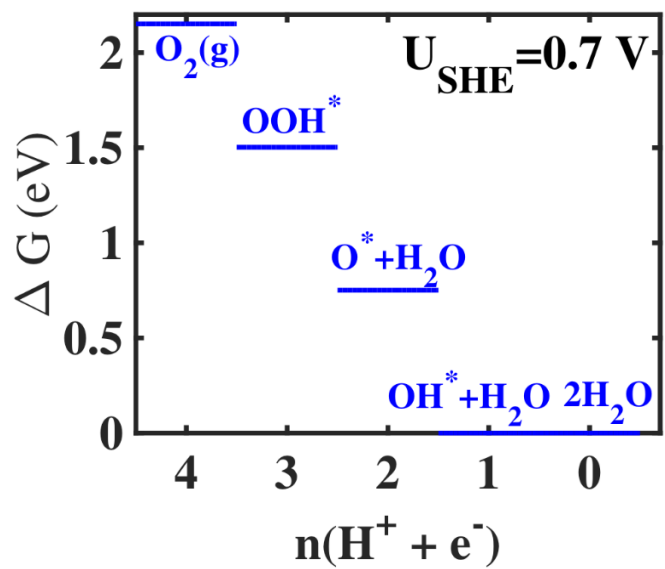

**Figure S23.** Free energy diagram plotted at the limiting potential on the Ni$_9$S$_8$ phase. The step involving the reduction of OH* to H₂O is the potential determining step and the free energy levels of these two intermediates line up at the limiting potential of 0.7 V.

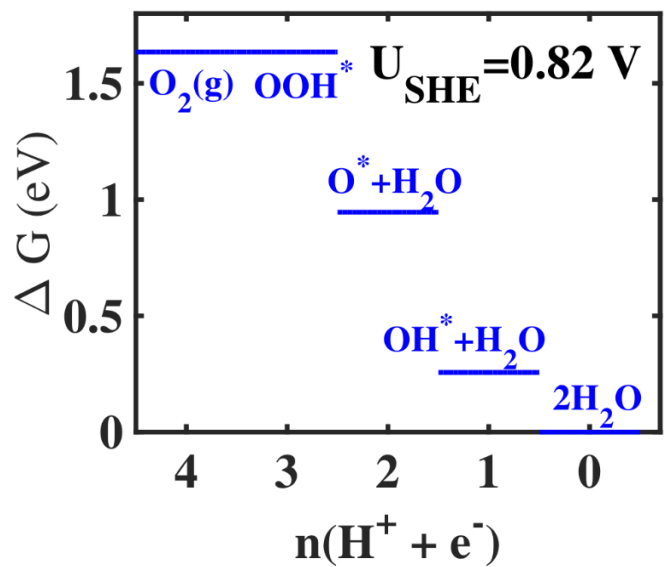

**Figure S24.** Free energy diagram plotted at the limiting potential on the $Ni_3S_2$ phase. The step involving the activation of $O_2$ as OOH* is the potential determining step and the free energy levels of these two intermediates line up at the limiting potential of 0.82 V.